\documentclass{article}
\usepackage[english]{babel}
\usepackage[utf8]{inputenc}
\usepackage{johd}
\usepackage{amsmath}
\usepackage{amsfonts}       
\usepackage{amssymb}        

\title{Deep Learning Assisted Outer Volume Removal for Highly-Accelerated Real-Time Dynamic MRI}

\author{Merve~G\"ulle$^{a}$$^{b}$, Sebastian~Weing\"artner$^{c}$$^{d}$, Mehmet~Ak\c{c}akaya$^{a}$$^{b}$$^{*}$ \\
        \small $^{a}$Department of Electrical and Computer Engineering, University of Minnesota, Minneapolis, MN, United States \\
        \small $^{b}$Center for Magnetic Resonance Research, University of Minnesota, Minneapolis, MN, United States \\
        \small $^{c}$Department of Imaging Physics, Delft University of Technology, Delft, the Netherlands \\
        \small $^{d}$HollandPTC, Delft, the Netherlands \\\\
        \small $^{*}$Corresponding author: Mehmet~Ak\c{c}akaya; \tt{akcakaya@umn.edu} \\
}
\date{}

\begin{document}
\maketitle

\begin{abstract} 
Real-time (RT) dynamic MRI plays a vital role in capturing rapid physiological processes, offering unique insights into organ motion and function. Among these applications, RT cine MRI is particularly important for functional assessment of the heart with high temporal resolution. RT imaging enables free-breathing, ungated imaging of cardiac motion, making it a crucial alternative for patients who cannot tolerate conventional breath-hold, ECG-gated acquisitions. However, achieving high acceleration rates in RT cine MRI is challenging due to aliasing artifacts from extra-cardiac tissues, particularly at high undersampling factors. In this study, we propose a novel outer volume removal (OVR) method to address this challenge by eliminating aliasing contributions from non-cardiac regions in a post-processing framework. Our approach estimates the outer volume signal for each timeframe using composite temporal images from time-interleaved undersampling patterns, which inherently contain pseudo-periodic ghosting artifacts. A deep learning (DL) model is trained to identify and remove these artifacts, producing a clean outer volume estimate that is subsequently subtracted from the corresponding k-space data. The final reconstruction is performed with a physics-driven DL (PD-DL) method trained using an OVR-specific loss function to restore high spatio-temporal resolution images. Experimental results show that the proposed method at high accelerations achieves image quality that is visually comparable to clinical baseline images, while outperforming conventional reconstruction techniques, both qualitatively and quantitatively. The proposed approach provides a practical and effective solution for artifact reduction in RT cine MRI without requiring acquisition modifications, offering a pathway to higher acceleration rates while preserving diagnostic quality.
\end{abstract}

\noindent\keywords{Real-time MRI; Dynamic MRI; Cine CMR; Ghosting artifacts; Outer volume removal; Unrolled networks; DL-based reconstruction}\\

\section{Introduction}\label{sec:introduction}

Real-time (RT) dynamic MRI facilitates visualization of physiological changes in real-time, using rapid snapshot acquisitions \citep{nayak2022}. These acquisitions are particularly useful for capturing non-periodic processes, where conventional gating or averaging methods may fail \citep{nayak2022}. Such acquisitions are important in diverse clinical applications, ranging from assessment of cardiac function \citep{wang2021fast,zhang2014real} to guidance of interventional procedures \citep{vanDenBosch2008, chubb2017development} to upper airway imaging \citep{Lingala2016, Zu2013} to musculoskeletal applications \citep{Kawchuk2015}. Among these, real-time cine cardiac MRI (CMR) is commonly used clinically as an alternative to breath-hold (BH) segmented cine CMR \citep{longere2023new,  wang2021fast}. The latter is considered the gold standard imaging technique for functional evaluation of the heart and is acquired with electrocardiogram (ECG) gating in BH to minimize motion artifacts and enhance image clarity \citep{ishida2009cardiac}. However, this method poses significant challenges for patients with arrhythmia, and for those who are unable to hold their breath, such as young children or individuals with respiratory issues \citep{roy2022free}, making the acquisition process difficult and often leading to suboptimal image quality \citep{rajiah2023cardiac}. For these patients, RT cine CMR is essential to allow free-breathing imaging without the need for ECG synchronization \citep{nayak2022, beer2010free}. RT cine CMR acquires dynamic images of the heart in succession using snapshot imaging, accommodating irregular heartbeats and breathing, which are critical in cases where traditional BH and ECG-gated protocols are not feasible \citep{UnterbergBuchwald2014}. High spatio-temporal resolution for RT cine CMR is crucial to accurately capture rapid cardiac motion and detailed anatomical structures \citep{Setser2000}.

Current RT CMR techniques often utilize parallel imaging \citep{kellman2001adaptive, breuer2005dynamic} and compressed sensing \citep{jung2009k, otazo2010combination, feng2011highly} to accelerate MRI scans. While these techniques have shown advancements in reducing scan times and improving image quality, they face challenges in achieving adequate spatio-temporal resolution for dynamic cardiac imaging due to the low acceleration rates \citep{tolouee2018motion, jung2009k, li2018real}. On the other hand, some studies employ spatio-temporal regularization techniques during the reconstruction process \citep{liu2020dynamic, feng2016xd, hauptmann2019real, hansen2012retrospective, demirel2023high, otazo2015low, pedersen2009k, tsao2003k}. These techniques reduce artifacts by leveraging the underlying spatial and temporal correlations but can introduce risks of temporal blurring, despite the critical importance of preserving sharp temporal features in dynamic cardiac imaging \citep{najeeb2020respiratory}. Thus, novel reconstruction techniques that enable high acceleration rates without requiring temporal regularization are desirable for RT CMR.

A major challenge in achieving higher acceleration rates for RT CMR using current techniques is related to the large amount of outer volume tissue that surrounds the heart \citep{smith2012reduced}. When imaging the heart, a large field of view (FOV), including areas such as the chest and back, is prescribed to avoid foldover. These outer volume tissues contain high-intensity fat signals, which contribute to high-intensity aliasing artifacts in the region of interest (ROI) encompassing the heart. At high acceleration rates, these present substantial challenges during reconstruction, potentially degrading image quality \citep{griswold2004field}. To address this challenge in broader cardiac MRI applications, one strategy involves the use of outer volume suppression (OVS) pulses or modules \citep{smith2012reduced, yang2018reduced, weingartner2018feasibility, wang2021high, luo2015combined, coristine2015combined}. These techniques aim to selectively suppress signals from surrounding extra-cardiac tissues that could otherwise degrade image quality. However, acquisitions that incorporate OVS modules have downsides, including signal recovery during imaging, high specific absorption rates (SAR), imperfect suppression, and importantly, disruptions to steady-state imaging, which limits its use in acquisitions such as cine CMR. As an alternative, view-sharing or keyhole techniques have been proposed in an attempt to capture moving parts in a larger ROI \citep{gomez2021clinical,fan2024ultra}. These techniques acquire k-space data with different sub-sampling patterns, and share information across dynamics, and have also been combined with outer volume estimation \citep{gomez2021clinical}. However, their adoption in clinical practice has been limited due to residual temporal artifacts and challenges in accurately estimating extra-cardiac volumes.

In this study, we take a different approach by proposing a reconstruction method that removes extra-cardiac tissue signals during the post-processing stage. This technique simplifies the reconstruction problem by significantly reducing aliasing artifacts within the ROI. Importantly, it is applicable to existing acquisition schemes and avoids issues associated with signal regrowth commonly encountered in OVS-based acquisitions. The proposed outer volume removal (OVR) method first estimates the outer volume signal for each timeframe using composite temporal images from time-interleaved shifted dynamic acquisitions. While these composite images tend to exhibit ghosting artifacts due to motion, a deep learning (DL) model is used to estimate and remove these artifacts. The estimated outer volume signal is subsequently removed from the corresponding k-space raw data of each timeframe. Finally, a physics-driven DL (PD-DL) network is trained with an OVR-specific loss function to reconstruct this high temporal resolution data in a frame-by-frame manner. Our results show that the proposed method outperforms conventional methods and PD-DL reconstruction without OVR both qualitatively and quantitatively. Our contributions are as follows:
\begin{itemize}
    \item We introduce a novel analytical characterization for the pseudo-periodic ghosting artifacts observed in low-temporal resolution composite images formed from time-interleaved shifted undersampling patterns, not considered in the literature before. 
    \item We propose a DL-based strategy to robustly estimate these motion-induced ghosting artifacts from composite images, which is used to generate a clean estimate of the outer volume signal that is subsequently removed from each individual timeframe.
    \item We propose a novel OVR-augmented loss function for training the PD-DL reconstruction network with improved fidelity, by promoting consistency between outputs corresponding to forward operators comprising coil sensitivities that span the whole FOV versus those that are constrained to the ROI.
    \item We conduct extensive experiments across different field strengths using both retrospective and prospective accelerated datasets and demonstrate that our method outperforms conventional and state-of-the-art acceleration methods without OVR qualitatively and quantitatively.
\end{itemize}

\section{Characterization of motion-induced ghosting in outer volume estimation}

\begin{figure}[!t]
  \includegraphics[width=\textwidth]{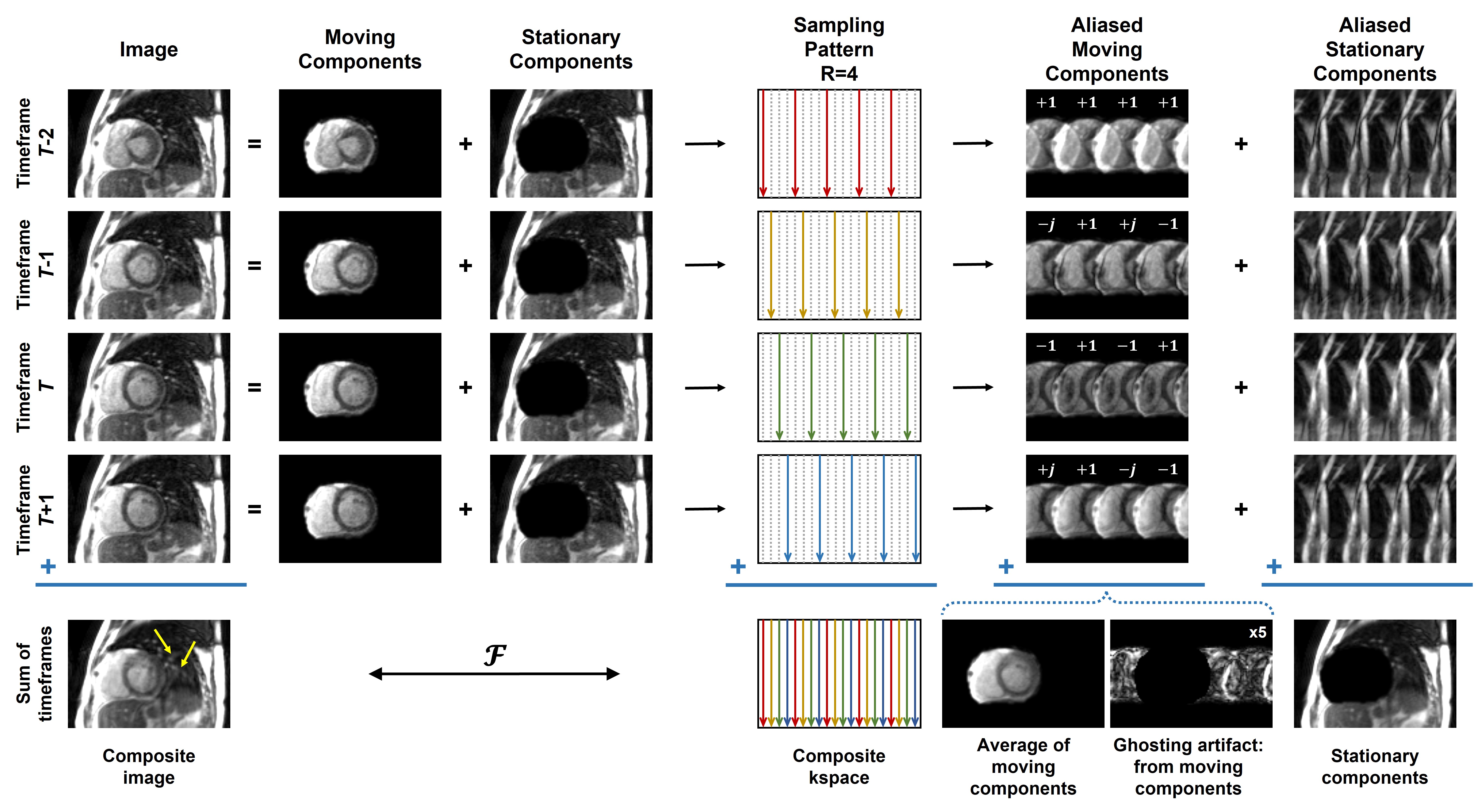}
  \caption{Illustration of the decomposition of a composite real-time cine cardiac MR image into various components. For simpler visualization, the heart is depicted as the only moving object, while surrounding tissues are treated as stationary, and $R = 4$ is used. Both the moving components and stationary components of each time-frame contribute to the composite image as foldovers with distinct modulation coefficients due to the shifted phase encoding lines across time-frames. The moving components add constructively at the true heart location, creating a temporally averaged representation, while their summation at the other foldover locations leads to a pseudo-periodic ghosting artifact due to cardiac motion between time-frames and the distinct modulation coefficients. Conversely, the stationary components add constructively at the central foldover while canceling out at the other foldover locations.}
  \label{fig: ghosting definition}
\end{figure}

Let $\mathbf{x}(t) \in \mathit{\mathbb{C}}^N$ represent the $t^{th}$ complex-valued timeframe of a dynamic MRI sequence, where $N$ is the number of pixels in the spatial plane, which is assumed to be 2D without loss of generality. The acquired k-space data, denoted as $\mathbf{y}^{\Omega_t}(t) \in \mathit{\mathbb{C}}^M$, corresponds to the measurements from k-space locations $\Omega_t$, with $M$ being the number of acquired k-space points per timeframe. Using a time-interleaved shifted equidistant or uniform undersampling pattern, as implemented in TGRAPPA \citep{breuer2005dynamic} or TSENSE \citep{kellman2001adaptive}, one can generate a fully sampled composite k-space or image with low temporal resolution by combining $R$ consecutive timeframes into a single merged dataset for acceleration rate $R$. Previous studies \citep{gomez2021clinical, blaimer2008sparse} have utilized this composite data to estimate the stationary background by masking the background image. However, residual temporal artifacts have remained a major challenge in these approaches.

In this study, we propose a novel analytical perspective on such composite data by treating the images from each timeframe as a combination of moving, $\mathbf{x}_\textrm{moving}(t)$, and stationary. $\mathbf{x}_\textrm{stationary}(t)$ components as:
\begin{equation}
    \mathbf{x}(t)=\mathbf{x}_\textrm{moving}(t)+\mathbf{x}_\textrm{stationary}(t).
\end{equation}

Fig.~\ref{fig: ghosting definition} illustrates this decomposition across different timeframes, where for simplicity, we depict the heart as the only moving object, while the surrounding tissues are considered stationary across these timeframes. For RT sequences, these assumptions approximately hold over the acquisition of $R$ subsequent frames, provided the temporal resolution is sufficiently small, since the respiratory motion is much slower compared to cardiac motion. 
In this scenario, each individual timeframe contributes an aliased image of the heart and the stationary background to the composite image. Importantly, due to the time-interleaved shifted pattern in the k-space acquisitions, each foldover of the aliased components has a distinct modulation phase. In the composite image, the side foldovers of the aliased background tend to cancel each other out, resulting in a stationary background. Conversely, the foldovers of the aliased heart images align at the true heart location, forming a temporally averaged representation of the heart at its correct position. Finally, other foldovers of the moving components add up to produce ghosting artifacts in the background due to differing modulation phases, which manifests as a pseudo-periodic pattern. 

Overall, this formulation allows us to decompose the composite image into the combination of temporarily averaged moving components $\mathbf{\overline{x}}_\textrm{moving}(t)$, a pseudo-periodic ghosting artifact $\mathbf{x}_\textrm{ghost}(t)$ due to the moving tissue components, and a stationary background $\mathbf{x}_\textrm{stationary}(t)$ as:
\begin{equation}
\mathbf{x}_{\textrm{com}}(t)=\underbrace{\mathbf{\overline{x}}_\textrm{moving}(t)+\mathbf{x}_\textrm{ghost}(t)}_\textrm{moving components}+\underbrace{\mathbf{x}_\textrm{background}(t)}_\textrm{stationary components}.
\label{Eq:com_img}
\end{equation}

\begin{figure}[!t]
  \includegraphics[width=\columnwidth]{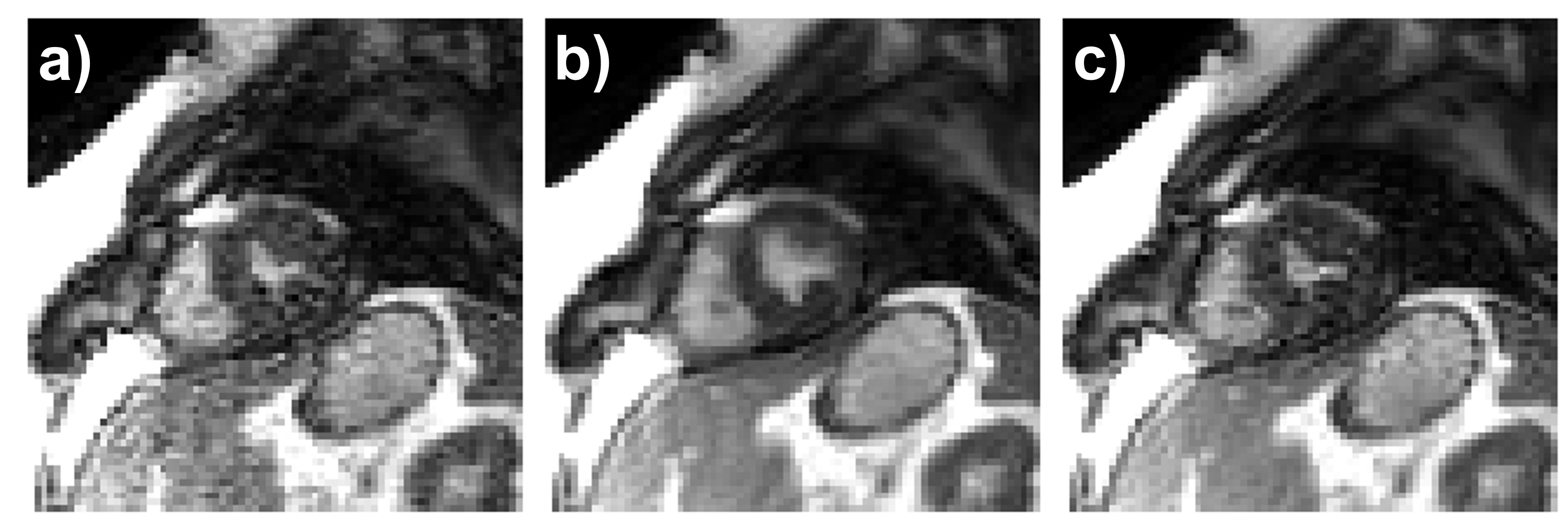}
  \centering
  \caption{Importance of accounting for ghosting artifacts during OVR. (a) TGRAPPA reconstruction at acceleration rate $R$ = 4. (b) CG-SENSE reconstruction of OVR k-space data at $R$ = 4, using composite coil images as the outer volume, without ghosting artifact correction. (c) CG-SENSE reconstruction of OVR k-space data at $R$ = 4, where the outer volume is estimated after ghosting artifact removal from composite images. The left column shows the baseline image during systole. The reconstruction of the same data in the middle column shows temporal blurring, effectively yielding a different cardiac phase due to the effect of the ghosting artifacts present in the data, while also exhibiting blurring artifacts. The right column matches the correct cardiac phase and shows high image fidelity.}
  \label{fig: OVR_comparison}
\end{figure}

Identifying and removing these ghosting artifacts is of paramount importance to maintain the temporal fidelity of the reconstructions after OVR. Fig.~\ref{fig: OVR_comparison} illustrates reconstructions with outer volume subtraction, where the outer volume is estimated from the composite images directly without accounting for the ghosting artifacts, and using our approach that identifies and removes these ghosting artifacts. The baseline data, shown in the left column, is selected from a systolic cardiac phase and presents the TGRAPPA reconstruction of the $R$ = 4 k-space data. In the middle column, a naive outer volume subtraction approach that uses the outer volume directly from the composite image is shown. The right column shows outer volume subtraction, where the ghosting artifacts are removed from the composite image prior to estimating the outer volume. The OVR k-space signal is calculated accordingly (explained in Sec.~\ref{sec: ghosting}) and reconstructed with CGSENSE. The reconstruction with the naive outer volume estimation loses temporal fidelity since the ghosting artifacts from other cardiac phases affect the final reconstruction. As a result, cardiac phase mismatch and significant blurring are observed. On the other hand, the outer volume estimation that explicitly considers ghosting artifacts yields a reconstruction that aligns with the correct cardiac phase and maintains superior temporal fidelity.

We note that in contrast to the simplified depiction in Fig.~\ref{fig: ghosting definition}, not only the heart but also other nearby tissues, such as the diaphragm, move rapidly enough to contribute to such ghosting artifacts. Consequently, while we use this simplified version in our preliminary work \citep{gulle2024robust} and for illustrative purposes here, our implementations in this study capture these three components without explicitly delineating boundaries for moving organs, as further described in Sec. \ref{sec: ghosting}.

\section{Methods}
\begin{figure}[!t]
  \includegraphics[width=\textwidth]{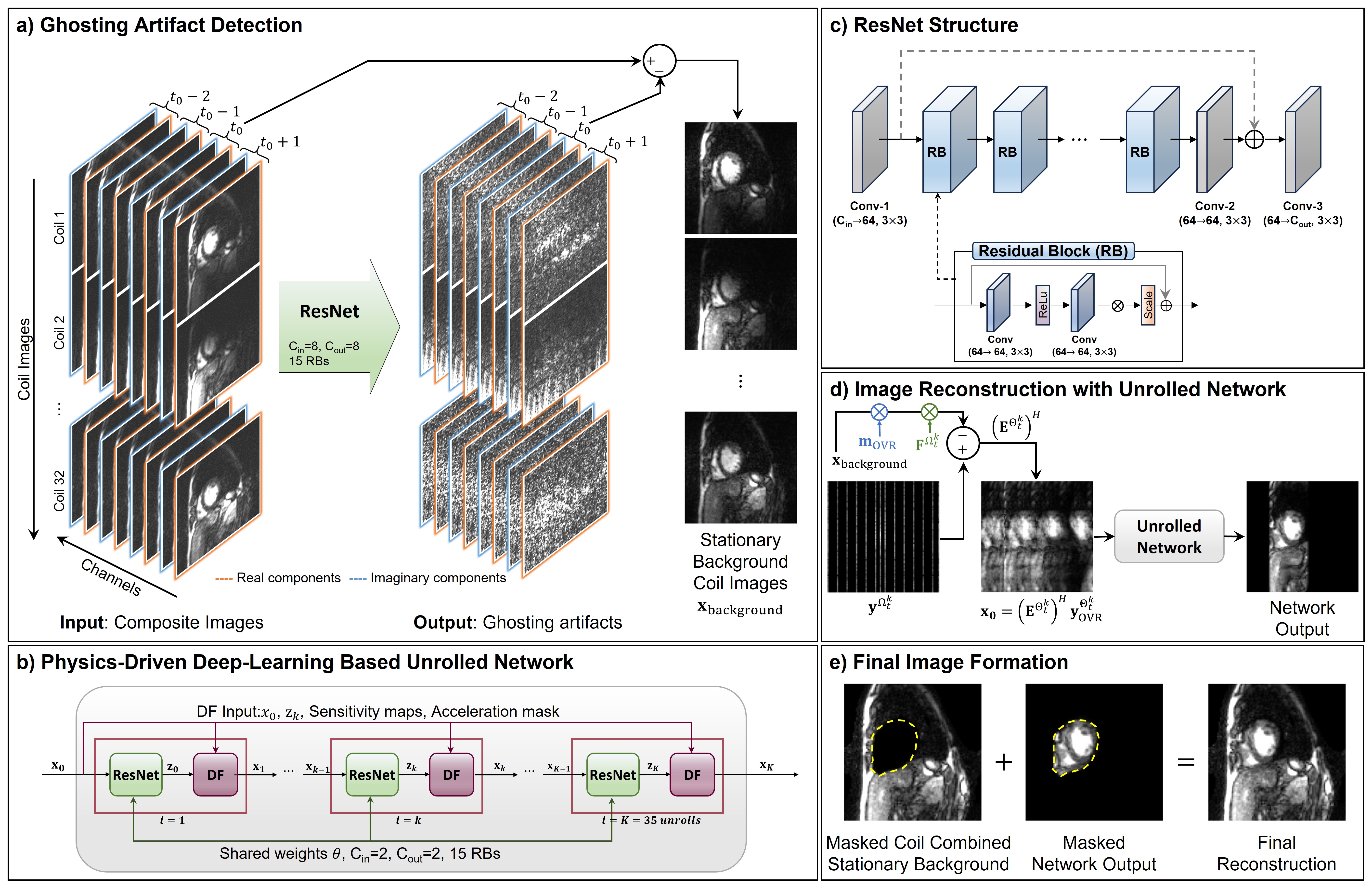}
  \caption{Proposed reconstruction pipeline: (a) DL-based ghosting detection: A ResNet with 15 residual blocks (RBs) takes four adjacent composite images, concatenating their real and imaginary components into eight input channels ($\textrm{C}_\textrm{in}=8$), and estimates the ghosting artifacts for the corresponding time frame $t_0$, producing eight output channels ($\textrm{C}_\textrm{out}=8$). The stationary background coil images, $\mathbf{x}_\textrm{background}$, are then obtained by subtracting the detected ghosting artifacts from the composite images of the target time frame. (b) Physics-driven DL-based unrolled network: The network consists of 35 unrolls, where each iteration block contains one ResNet and one data fidelity (DF) block. The ResNets ($\textrm{C}_\textrm{in}=2$, $\textrm{C}_\textrm{out}=2$, 15 residual blocks) perform the proximal operation for the regularizer on the previous DF block output, and share the same network weights ($\theta$). The DF block takes the zerofilled image ($\mathbf{x}_\textrm{0}$), the last ResNet output ($\mathbf{z}_\textrm{k}$), sensitivity maps, and the acceleration mask, producing a data-consistent image as output using conjugate gradient. (c) The ResNet structure is used in both the ghosting detection network and the proximal operators of the unrolled network. (d) Outer volume subtraction and PD-DL reconstruction: The background image is masked using the OVR mask ($\mathbf{m}_\textrm{OVR}$) and transformed into k-space via the Fourier transform ($\mathcal{F}^{\Omega_t}$), then subtracted from the acquired signal ($\mathbf{y}^{\Omega_t}$). The OVR k-space signal $\mathbf{y}_\textrm{OVR}^{\Omega_t}$ is mapped back to the image domain using the adjoint encoding operator $\big ( \mathbf{E}^{\Omega_t} \big ) ^H$ and fed into the unrolled network. As expected, the network output exhibits minimal background signal. (e) Final image formation by combining the masked background with the reconstruction.}
  \label{fig: method_fig}
\end{figure}

\subsection{DL-based ghosting detection and outer volume removal (OVR)}\label{sec: ghosting}
Let $\mathbf{x}_\textrm{com}(t_0)$ be the composite image of time $t_0$. To estimate the background of this acquisition $\mathbf{x}_\textrm{background}(t_0)$, we first propose a DL-based technique to estimate the motion-related pseudo-periodic ghosting artifacts in the composite images illustrated in Fig.~\ref{fig: method_fig}a. The network takes 4 adjacent composite images in channel-wise concatenated form, represented as $\mathbf{x}_{\textrm{com}}^{\textrm{concat}}(t_0)\triangleq \{\mathbf{x}_\textrm{com}(\tau)\}^{t_0+1}_{\tau=t_0-2}$, covering a wider temporal window to leverage the correlation over timeframes. The network outputs the corresponding ghosting artifacts of each composite image in the output, again in a channel-wise concatenated form $\mathbf{x}_{\textrm{ghost}}^{\textrm{concat}}(t)\triangleq \{\mathbf{x}_\textrm{ghost}(\tau)\}^{t_0+1}_{\tau=t_0-2}$. We note that all four images are used for loss calculation, but only the ghosting estimate at $t_0$ is used for subsequent OVR for the corresponding timeframe.

The network is trained in a supervised manner by minimizing a normalized $\ell_2$ loss between the network output and the reference ghosting artifacts
\begin{equation} 
    \min _{\boldsymbol{\theta}} {\mathbb E} \bigg[
    {\cal L}_{\textrm{normalized-}\ell_2} \big(\mathbf{x}_{\textrm{ghost-ref}}^{\textrm{concat}}(t),  f_{\boldsymbol{\theta}}\left(\mathbf{x}_{\textrm{com}}^{\textrm{concat}}(t)\right)\big) \bigg],
    \label{Eq:cost}
\end{equation}
where $\mathbf{x}_{\textrm{ghost-ref}}^{\textrm{concat}}(t)$ is the concatenated reference ghosting images, the generation of which is detailed in Sec. \ref{sec: ghosting network}, and $f_{\boldsymbol{\theta}}\left( \cdot \right)$ is the ghosting detection network with trainable parameters $\boldsymbol{\theta}$. After the estimation of these ghosting artifacts, their contribution is subtracted from the corresponding composite images to obtain clean outer volume background images:
\begin{equation} \label{eq:4}
    \mathbf{x}_\textrm{background}(t_0) \approx \mathbf{x}_\textrm{com}(t_0) - f_{\boldsymbol{\theta}}\left( \mathbf{x_i}_{\textrm{com}}^{\textrm{concat}}(t_0) \right)\big|_{t=t_0}
\end{equation}
where $f_{\boldsymbol{\theta}}\left( \mathbf{x_i}_{\textrm{com}}^{\textrm{concat}}(t_0) \right)\big|_{t=t_0}$ refers to the estimated ghosting artifact of the time of interest $t_0$ over the four timeframes output by the neural network.

Finally, once the background signal is estimated according to \eqref{eq:4}, the last step for OVR is to subtract it from the raw k-space data without interfering with the signal around the heart. To this end, an additional neural network was trained to predict heart boundaries from coil-combined composite images. This was used to generate a mask, $\mathbf{m}_\textrm{OVR}$, that outlines the outer volume to be removed. The outer volume was defined as everything outside the heart boundaries. These boundaries were specified using a rectangle in the phase-encode direction, spanning the full frequency-encode direction (since no undersampling is performed in the latter), as illustrated in Fig.~\ref{fig: method_fig}d. Subsequently, the OVR k-space was generated from the acquired data $\mathbf{y}^{\Omega_t}$ as:
\begin{equation}
    \mathbf{y}_\textrm{OVR}^{\Omega_t} = \mathbf{y}^{\Omega_t} - \mathcal{F}^{\Omega_t}\{\mathbf{m}_\textrm{OVR} \cdot \mathbf{x}_\textrm{background}  \},
\end{equation}
where $\mathcal{F}^{\Omega_t}$ denotes the Fourier transform operator undersampled at ${\Omega_t}$ k-space locations. The DL-based generation of OVR masks and network details are described in Sec. \ref{sec: OVR mask}.

\subsection{PD-DL MRI reconstruction}\label{sec: PDDL recon}
Let $\mathbf{y}_\textrm{OVR}^{\Omega_t}$ be the outer volume-removed k-space signal with $\Omega_t$ as the undersampling pattern for timeframe $t$. PD-DL reconstruction solves a regularized least-squares problem:
\begin{equation} 
    \arg \min _{\mathbf{x}_\textrm{OVR}}\left\|\mathbf{y}_\textrm{OVR}^{\Omega_t}-\mathbf{E}^{\Omega_t} \mathbf{x}_\textrm{OVR}\right\|_2^2+\mathcal{R}\left(\mathbf{x}_\textrm{OVR}\right),
    \label{Frwd_cost}
\end{equation}
where $\mathbf{E}^{\Omega_t}$ is the multi-coil encoding operator with a sub-sampled Fourier operator that samples $\Omega_t$ and coils sensitivity maps, and $\mathbf{x}_\textrm{OVR}$ is the target image with the outer volume excluded. The objective function consists of two terms: a data fidelity (DF) term that ensures consistency with the acquired k-space data and a regularization term $\mathcal{R}(\cdot)$ that imposes prior constraints on the reconstruction.

A common approach in PD-DL reconstruction is to unroll an iterative algorithm \citep{hammernik2023physics, zhu2023physics, ramzi2022nc, yaman2020self, aggarwal2018modl, hammernik2018learning, schlemper2017deep, sun2016deep} to solve (\ref{Frwd_cost}) over a fixed number of iterations. The proximal operator for $\mathcal{R}(\cdot)$ is implicitly modeled using a neural network, while the DF term weights are learned during end-to-end training. The unrolled network structure is shown in Fig.~\ref{fig: method_fig}b-c, and its implementation details are provided in Sec.~\ref{sec: PDDL_description}.

\subsection{Training of the PD-DL network and effect of sensitivity maps in OVR}
Noting that there is no fully-sampled ground-truth available for RT acquisitions, an unsupervised learning method is necessary for training the associated PD-DL reconstruction network \citep{akcakaya2022unsupervised}. To this end, we adopt a self-supervised learning approach known as Self-Supervised learning via Data Undersampling (SSDU) \citep{yaman2020self, yaman2022multi}, which eliminates the need for fully sampled reference data. Specifically, the multi-mask version of SSDU partitions the acquired k-space data locations, $\Omega_t$, into multiple disjoint sets $(\Theta_t^k, \Lambda_t^k)_{k=1}^K$. The first subsets, $\Theta_t^k$, are used for the DF term during network training, while the second subsets, $\Lambda_t^k$, are used to define the self-supervised loss. The resulting multi-mask SSDU loss function is given by:
\begin{equation}
    \min _{\boldsymbol{\gamma}} {\mathbb E} \Bigg[
    \frac1K \sum_{k=1}^K \mathcal{L}\biggl(\mathbf{y}_\textrm{OVR}^{\Lambda_t^k}, \mathbf{E}^{\Lambda_t^k}\Bigl(g_{\boldsymbol{\gamma}}\bigl(\mathbf{y}_\textrm{OVR}^{\Theta_t^k}, \mathbf{E}^{\Theta_t^k} \bigl)\Bigl)\biggl) \Bigg],
    \label{SSDU_cost}
\end{equation} 
where $g_{\boldsymbol{\gamma}}\bigl(\mathbf{y}_\textrm{OVR}^{\Theta_t^k}, \mathbf{E}^{\Theta_t^k} \bigl)$ denotes the output of the PD-DL network parameterized by $\boldsymbol{\gamma}$, using the input measurements $\mathbf{y}_\textrm{OVR}^{\Theta_t^k}$ and encoding operator $\mathbf{E}^{\Theta_t^k}$. $K$ is the number of SSDU masks, and $\mathcal{L}(\cdot, \cdot)$ is the normalized $l_1$-$l_2$ loss function \citep{Knoll_SPM}.

\begin{figure}[!t]
  \includegraphics[width=\columnwidth]{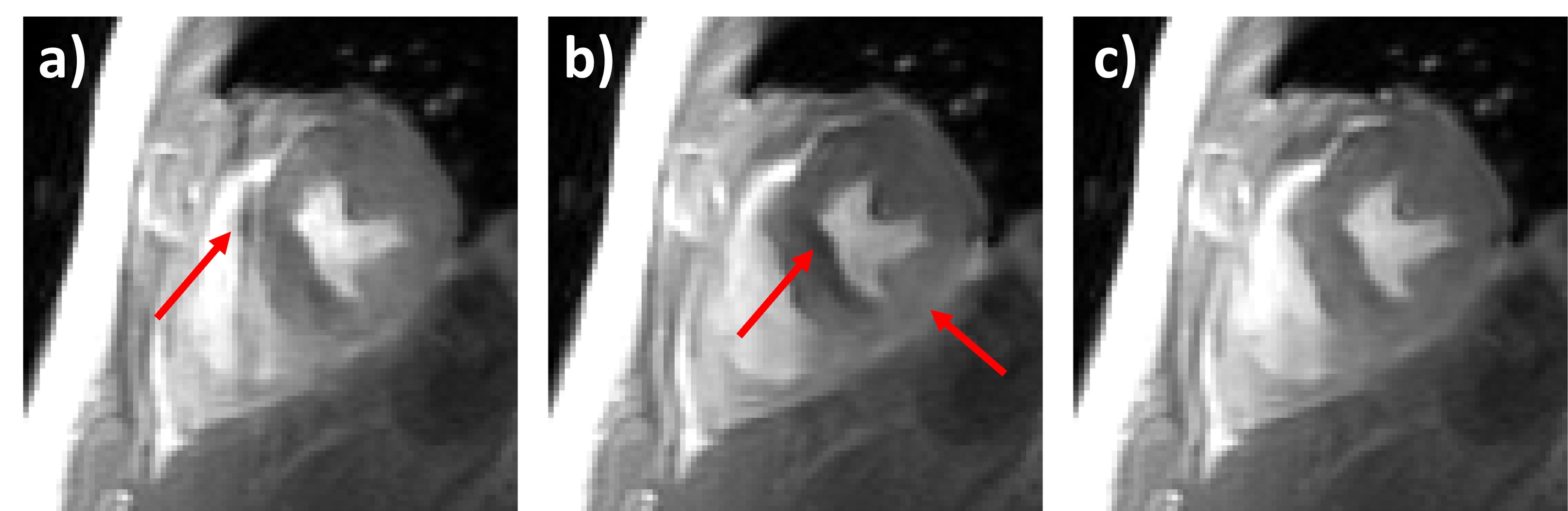}
  \centering
  \caption{PD-DL reconstruction of the OVR k-space using (a) masked sensitivity maps, (b) full sensitivity maps, and (c) full sensitivity maps with consistency through the proposed loss function in \eqref{full_cost}. The left column shows artifacts, while the middle column exhibits signal loss. The proposed consistency mechanism effectively mitigates both issues, as demonstrated in the right column.}
  \label{fig: Smaps_comparison}
\end{figure}

While this loss may be used naively, additional considerations apply to our OVR scenario. If the outer volume is indeed perfectly removed from $\mathbf{y}_\textrm{OVR}^{\Lambda_t^k}$, then the corresponding pixels in $\mathbf{x}_\textrm{OVR}$ are zero. Note that this is a stronger condition than sparsity in compressed sensing, since the location of the zero coefficients is specified as well. This can be enforced either via masking in the image domain, or in the context of PD-DL, more appropriately through setting the coil sensitivities in the outer volume region to zero in ${\bf E}^{\Lambda_t^k}$ in order to not to affect the learning of the proximal operator. The latter can also be viewed as inverting a lower rank system in the context of SENSE \citep{pruessmann1999sense}. We refer to these sensitivity maps as ``masked'', conversely, we refer to the unaltered maps as ``full'' sensitivity maps. 

However, this represents an idealized scenario, as some residual outer volume signal will remain in k-space. If masked sensitivity maps are used, any residual signal must be mapped into the retained pixels of $\mathbf{x}_\textrm{OVR}$, potentially introducing artifacts into the ROI, as illustrated in Fig.~\ref{fig: Smaps_comparison}a. Conversely, using full sensitivity maps results in a more challenging reconstruction problem and carries the risk of redistributing the ROI signal into the outer volume, leading to potential signal loss within the ROI, as illustrated in Fig.~\ref{fig: Smaps_comparison}b.

Thus, it is beneficial to combine the strengths of both approaches. To this end, we propose a novel loss function for training a PD-DL network in the OVR scenario. As noted, using masked sensitivities carries an inherent risk of residual outer volume signal within the ROI, which cannot be reversed. Therefore, we opt to use full sensitivity maps for the final reconstruction.
Nonetheless, the masked sensitivity maps are used as an additional term during training to ensure consistency in the ROI signal. In particular, we first train a PD-DL network for OVR k-space data using masked sensitivity maps with the loss function in \eqref{SSDU_cost}. Let $\mathbf{x}_\textrm{OVR}^{masked}$ denote the reconstructed images obtained from this network.
To avoid issues related to signal loss by naively using the full sensitivity maps, we introduce a loss function term that encourages consistency between reconstructions obtained with masked and full sensitivity maps within the ROI:
\begin{align}
    \min _{\boldsymbol{\gamma}} {\mathbb E} \Bigg[
    \frac1K &\sum_{k=1}^K \mathcal{L}\biggl(\mathbf{y}_\textrm{OVR}^{\Lambda_t^k}, \mathbf{E}^{\Lambda_t^k}\Bigl(g_{\boldsymbol{\gamma}}\bigl(\mathbf{y}_\textrm{OVR}^{\Theta_t^k}, \mathbf{E}^{\Theta_t^k} \bigl)\Bigl)\biggl) \Bigg] \nonumber \\
    &+ \lambda \mathcal{L}\Bigl(\mathbf{x}_\textrm{OVR}^{masked}, g_{\boldsymbol{\gamma}}\bigl(\mathbf{y}_\textrm{OVR}^{\Omega_t}, \mathbf{E}^{\Omega_t} \bigl) \cdot \mathbf{m}_\textrm{OVR} \Bigl),
    \label{full_cost}
\end{align} 
where $\lambda$ is a weight term, and the multi-coil operator $\mathbf{E}^{\Theta_t}$ are defined with full sensitivity maps. 
This loss function penalizes discrepancies between the two reconstructions, guiding the network to preserve the signal within the ROI, while allowing any residual outer volume signal to be mapped outside the ROI, thereby suppressing artifacts. As shown in Fig.~\ref{fig: Smaps_comparison}c, the updated network effectively eliminates both unwanted artifacts and signal loss in the ROI. 

\section{Experiments and implementation details}
\subsection{Imaging experiments} 
\subsubsection{Retrospectively accelerated datasets}
Data were acquired from 13 subjects in the left-ventricular short-axis using a bSSFP sequence with an acceleration factor of $R = 4$ and a Cartesian time-interleaved shifted uniform undersampling pattern \citep{breuer2005dynamic}. The imaging parameters were: field of view (FOV) = 360$\times$270 mm², acquired spatial resolution = 2.25$\times$2.93 mm² (reconstructed to 2.25$\times$2.25 mm²), slice thickness = 8 mm (11 slices), partial Fourier (PF) = 6/8, asymmetric echo = 20\%, echo time (TE)/repetition time (TR) = 1ms/2.34ms, leading to 17 phase encode (PE) lines per timeframe. To simulate higher acceleration, these datasets were retrospectively undersampled with an acceleration factor of $R=8$ by selecting every 8\textsuperscript{th} line while retaining the $k_y$ line nearest to the center of k-space for each timeframe, leading to 9 PE lines per timeframe. Note that the extra central line is retained since the center of the k-space contains the majority of the signal, which can be missed almost entirely at this high acceleration rate for certain timeframes. Practically, the extra line would lead to a mild change of 1 TR ($\sim$2ms) in temporal resolution.

\subsubsection{Prospectively accelerated datasets} \label{sec: prospective experiments}
RT cine MRI data were collected at 3T using a gradient-echo (GRE) sequence with time-interleaved Cartesian sampling in the left-ventricular short-axis. The imaging parameters were: acquired spatial resolution = 1.70$\times$2.27 mm², FOV = 300$\times$300 mm², slice thickness = 5 mm across 12 slices, PF = 6/8, asymmetric echo = 20\%, TE = 2.4 ms, and TR = 4.7 ms. Prospectively accelerated RT cine data were acquired from 18 subjects (12 for training, 6 for testing) at an acceleration rate of $R=8$, with a temporal resolution of 61 ms by acquiring 13 PE lines per timeframe. Additionally, 9 subjects were scanned with an acceleration rate of $R=4$ and a temporal resolution of 113 ms (24 PE lines per timeframe) to train the ghosting artifact detection network, as detailed in Sec. \ref{sec: ghosting network}.

Finally, for comparison, we also acquired BH ECG-gated segmented cine images on the same subjects. Data were collected from the same short-axis LV slices as the prospective $R=8$ scans, with a slice thickness of 5 mm across 12 slices per subject. The imaging parameters were the same as RT cine, except that temporal resolution was 47 ms, and no PF or in-plane acceleration was used.

\subsection{Implementation details} 
\subsubsection{Ghosting artifact detection network} \label{sec: ghosting network}
Two separate networks were trained for detecting ghosting artifacts, one for the bSSFP (retrospective) data and one for the GRE (prospective) data. The network architecture (Fig.~\ref{fig: method_fig}a) was based on a ResNet (Fig.~\ref{fig: method_fig}c), consisting of input and output convolutional layers and 15 residual blocks (RBs) \citep{yaman2022multi}. Each RB contains two convolutional layers, where the first is followed by a rectified linear unit (ReLU) and the second by a constant multiplication layer. All layers use a $3\times 3$ kernel size with 64 channels. Its input and output had 8 channels, processing the real and imaginary components of 2D images from 4 timeframes, as described in Sec. \ref{sec: ghosting}.

The network for retrospective data was trained on 352 slices from 9 subjects, using a learning rate (LR) of 10\textsuperscript{-3} over 250 epochs. Data acquired at an acceleration rate of $R = 4$ were retrospectively undersampled to $R = 8$, from which composite images were created using 8 subsequent timeframes. Reference ghosting artifact images were generated for each coil by subtracting TGRAPPA \citep{breuer2005dynamic} reconstructions of the $R = 4$ data from the composite images. For prospective data, the network was trained on 500 slices from 9 subjects with a learning rate of 10\textsuperscript{-3} for 250 epochs. Since no reference ghosting artifacts were available for the $R=8$ data, the auxiliary data acquired at $R = 4$ were used with retrospective undersampling. The ghosting reference was generated using the difference between composite images formed with $R=8$ data and TGRAPPA reconstructed $R=4$ data on a coil-by-coil basis. In both retrospective and prospective acceleration setups, the training was conducted by minimizing (\ref{Eq:cost}).

\subsubsection{OVR mask prediction network} \label{sec: OVR mask}
To estimate ${\bf m}_\textrm{OVR}$ from coil-combined composite images, we trained a U-Net \citep{ronneberger2015u} to predict heart boundaries. The network was trained using 1200 slices from 12 subjects with the binary cross-entropy loss, LR = 10\textsuperscript{-3} for 1000 epochs. Its input was a single coil-combined composite image, and its output was the corresponding binary mask of the same size.

\subsubsection{PD-DL reconstruction network} \label{sec: PDDL_description}
For both retrospective and prospective acceleration, we performed all trainings (for $\mathbf{x}_\textrm{OVR}^{masked}$ and $\mathbf{x}_\textrm{OVR}$) using the PD-DL network described in \cite{yaman2020selfconf, DemirelSIIM, demirel202120,hosseini2020dense} with 35 unrolls. The proximal operator for the regularizer was implemented using a ResNet \citep{yaman2020self,gu2022revisiting}, with shared parameters across the unrolls \citep{aggarwal2018modl}. The input to the unrolled network consisted of zero-filled images generated from outer volume-removed k-space data. We first used masked sensitivity maps and optimized the normalized $\ell_1$-$\ell_2$ loss function from Equation \ref{SSDU_cost} to train and generate $\mathbf{x}_\textrm{OVR}^{masked}$ images. For the final reconstruction, we used full sensitivity maps and a normalized $\ell_1$-$\ell_2$ loss with an additional $\mathbf{x}_\textrm{OVR}^{masked}$ feedback term, as described in (\ref{full_cost}), with $\lambda=.02$. 

For retrospective acceleration, both networks were trained on 352 slices from 9 subjects, and $K=3$ masks were used to implement the multi-mask SSDU algorithm \citep{yaman2022multi} using LR = 2$\cdot$10\textsuperscript{-4} over 200 epochs. The final reconstruction network was tested on 11 slices and 100 timeframes from 3 subjects, with data retrospectively undersampled from $R=4$ to $R=8$.

For prospective acceleration, both networks were trained on 480 slices from 12 subjects, acquired at $R=8$, and $K=3$ masks were used in multi-mask SSDU, LR = 2$\cdot$10\textsuperscript{-4} over 200 epochs. The final reconstruction network was tested on 10 slices and 80 timeframes from 6 subjects. 

\subsubsection{Comparison methods}
With retrospective acceleration, TGRAPPA \citep{breuer2005dynamic} at $R=4$ was used as baseline, while we note this is not a true reference due to noise amplification from parallel imaging reconstruction. For prospective acceleration, separate BH ECG-gated segmented cine acquisitions were used as the baseline. Note, since this is a different acquisition, the closest cardiac phase from the BH acquisition is used for display purposes. Furthermore, in all cases, for $R=8$, comparisons were made using TGRAPPA as a conventional benchmark, and the same PD-DL network in Sec. \ref{sec: PDDL_description} without OVR as the fair DL method comparison. For a fair comparison, this non-OVR PD-DL network was trained with the same slices as the proposed OVR PD-DL, but using the standard non-OVR loss function in \eqref{SSDU_cost}. Notably, all methods presented here are implemented without temporal regularization, ensuring high temporal resolution and a fair comparison.

\subsection{Cardiac function analysis}
Following the reconstructions, cardiac analysis was performed to compare the results of the proposed PD-DL with OVR and baseline reconstructions, TGRAPPA with an acceleration factor of $R=4$ for retrospective data, and BH cine for prospective data. Left ventricular (LV) functions were quantified using Segment v4.0 R12067b (Medviso, segment.heiberg.se) \citep{heiberg2010design}. The DL-based segmentation tool was used in the software, with manual adjustments to precisely delineate the boundaries of the myocardium to the blood. LV measurements were reported for 3 test subjects in the retrospective data and 6 test subjects in the prospective data. These data were first tested for Gaussianity using a Shapiro-Wilk test. Subsequently, the statistical differences in LV measurements were assessed using a paired t-test in the prospective study. A P-value $<$.05 was considered significant. Due to the limited number of test subjects, a t-test was not performed for the retrospective study.

\section{Results}
\subsection{Retrospectively accelerated datasets}
\subsubsection{Ghosting artifact detection}
\begin{figure}[tb]
    \centerline{\includegraphics[width=\columnwidth]{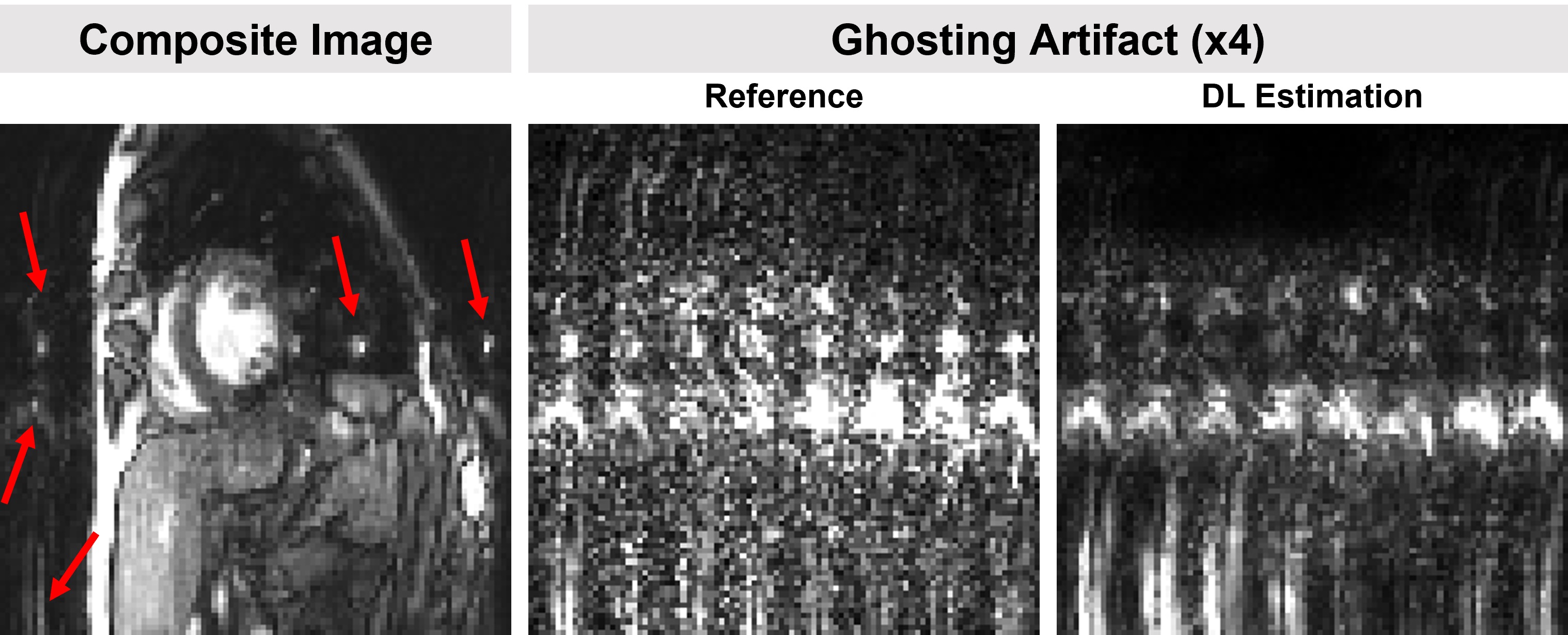}}
    \caption{Detection of the ghosting artifact of a composite RT cine image (retrospective acceleration $R=8$). The composite image (left) contains both static and dynamic structures, with motion-induced artifacts marked by red arrows, and the reference ghosting map (middle) is generated using TGRAPPA at $R=4$. The proposed method (right) effectively isolates ghosting artifacts, capturing their pseudo-periodic patterns. We also note that our method benefits from higher-SNR composite images, compared to the noisier reference labels from TGRAPPA ($R=4$), which suffer from spatially varying g-factor noise. As a result, the DL-based estimation represents a denoised version of the reference ghosting map, as expected.
    }
    \label{fig: ghosting_retro}
\end{figure}

Fig.~\ref{fig: ghosting_retro} depicts a representative composite image, the reference ghosting artifact as described in Sec. \ref{sec: ghosting network}, and the output of the DL-based ghosting estimation network. The composite image (left) contains both static and dynamic structures, where motion-related ghosting artifacts are highlighted with red arrows. The proposed DL model effectively isolates ghosting artifacts (right) similar to the reference labels (middle). The extracted ghosting patterns exhibit pseudo-periodic structures, demonstrating the model’s ability to capture and remove these artifacts. Additionally, the reference labels appear noisier due to their generation using TGRAPPA at an acceleration factor of $R=4$, which introduces spatially varying g-factor noise \citep{breuer2005dynamic}. In contrast, our method benefits from higher-SNR composite images as input. 

\subsubsection{RT cine MRI reconstructions}
\begin{figure}[!t]
    \centerline{\includegraphics[width=\columnwidth]{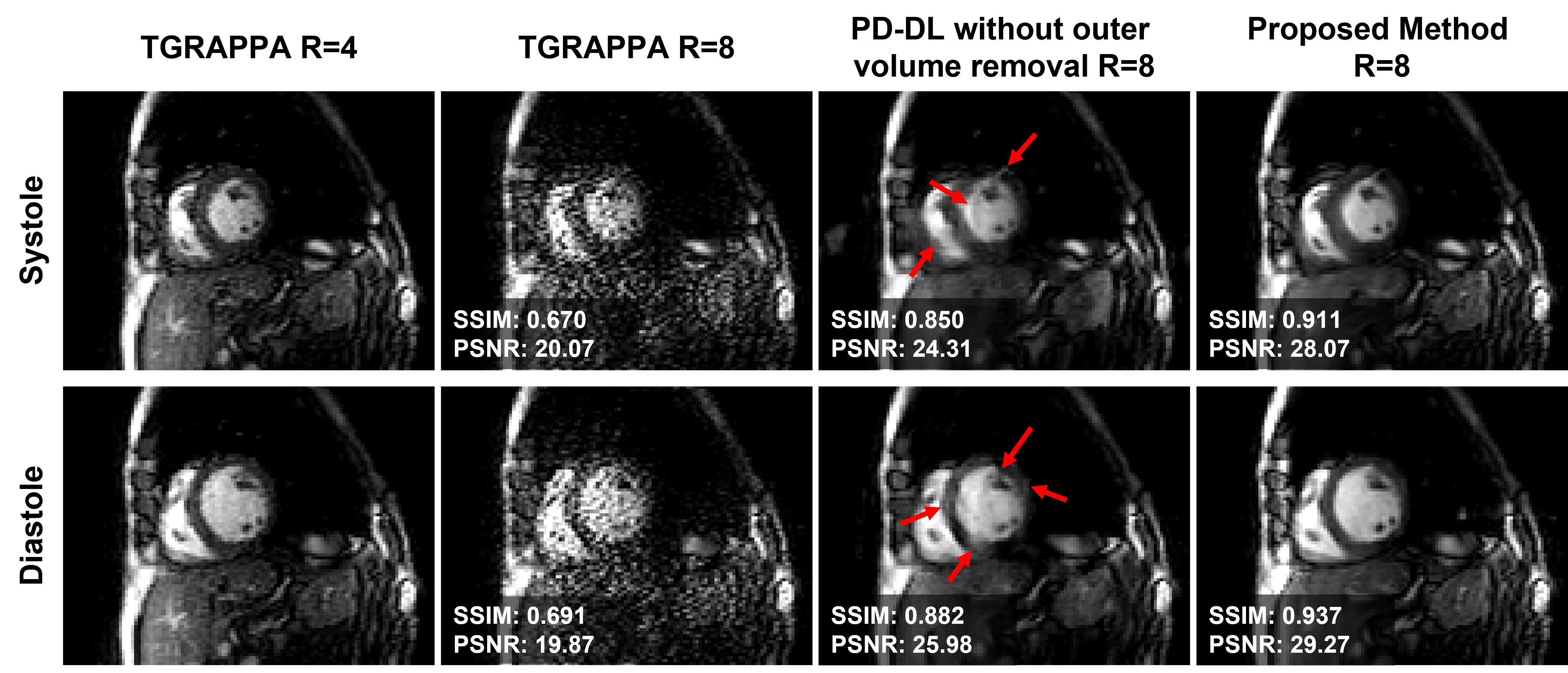}}
    \caption{Comparison of reconstructed images from a retrospectively $R$ = 8 accelerated RT cine dataset at systolic and diastolic phases. The first and second columns show TGRAPPA at baseline $R$ = 4, and at $R$ = 8, respectively. The latter exhibits severe artifacts. The PD-DL reconstruction without OVR (third column) reduces some artifacts but introduces blurring and signal loss, particularly at the myocardium and papillary muscles (red arrows). The proposed OVR-based method (last column) effectively mitigates these issues, restoring anatomical clarity and preserving the myocardium-blood interface. The resulting image quality is on par with TGRAPPA at $R=4$, demonstrating the impact of OVR in improving high-acceleration RT imaging.}
    \label{fig: retro_recon}
\end{figure}

Fig.~\ref{fig: retro_recon} shows two representative timeframes from the systolic and diastolic cardiac phases of a retrospectively $R$ = 8 accelerated RT cine dataset. The comparison includes the conventional TGRAPPA method for both $R$ = 4 (first column, baseline) and $R$ = 8 (second column), a conventional PD-DL technique without OVR (third column), and the proposed OVR-based method (rightmost column). TGRAPPA at $R$ = 8 suffers from substantial residual artifacts and very low signal-to-noise ratio (SNR). The PD-DL reconstruction without OVR reduces noise, but exhibits visible artifacts and blurring in the myocardium and papillary muscles (red arrows). This indicates that PD-DL alone struggles to maintain image clarity at high acceleration rates. On the other hand, the proposed method substantially mitigates these artifacts and provides a clearer depiction of anatomical structures, such as the myocardium-blood interface, with image quality comparable to TGRAPPA at baseline $R$ = 4. Quantitative evaluations of PSNR and SSIM values, relative to the baseline $R=4$ reconstruction, further support the improvements made by the proposed method. We emphasize that these metrics are imperfect, as they are influenced by the g-factor noise amplification in the TGRAPPA baseline $R$ = 4 reconstruction. 

\subsubsection{Quantitative cardiac function analysis}
Finally, the quantitative LV analysis comparing our proposed reconstruction at $R$ = 8 with the baseline $R$ = 4 reconstructions on the test dataset of 3 subjects is presented in Table~\ref{tab1}. The results show that the LV measurements from the proposed method are in close agreement with those from the baseline reconstructions, highlighting the reliability and accuracy of our approach. 

\begin{table}[H]
\centering
\caption{\label{tab1} LV analysis for retrospectively accelerated data. EDV: end diastolic volume; ESV: end systolic volume; EF: ejection fraction; SV: stroke volume; EDM: end diastolic mass; ESM: end systolic mass. Results are given as a mean, with the standard deviation in parentheses.}
\begin{tabular}{lll}
\hline
Metric & Proposed Method ($R$ = 8) & TGRAPPA ($R$ = 4) (Baseline) \\
\hline
EDV (mL) & 124.7 (21) & 120.8 (29) \\
ESV (mL) & 56.0 (13)  & 42.2 (15)  \\
EF (\%)  & 55.0 (6)   & 65.4 (7)   \\
SV (mL)  & 68.7 (12)  & 78.5 (18)  \\
EDM (g)  & 125.0 (18) & 109.8 (25) \\
ESM (g)  & 137.7 (30) & 105.7 (24) \\
\hline
\end{tabular}
\end{table}

\subsection{Prospectively accelerated datasets}
\subsubsection{Ghosting artifact detection}
\begin{figure}[!t]
    \centerline{\includegraphics[width=\columnwidth]{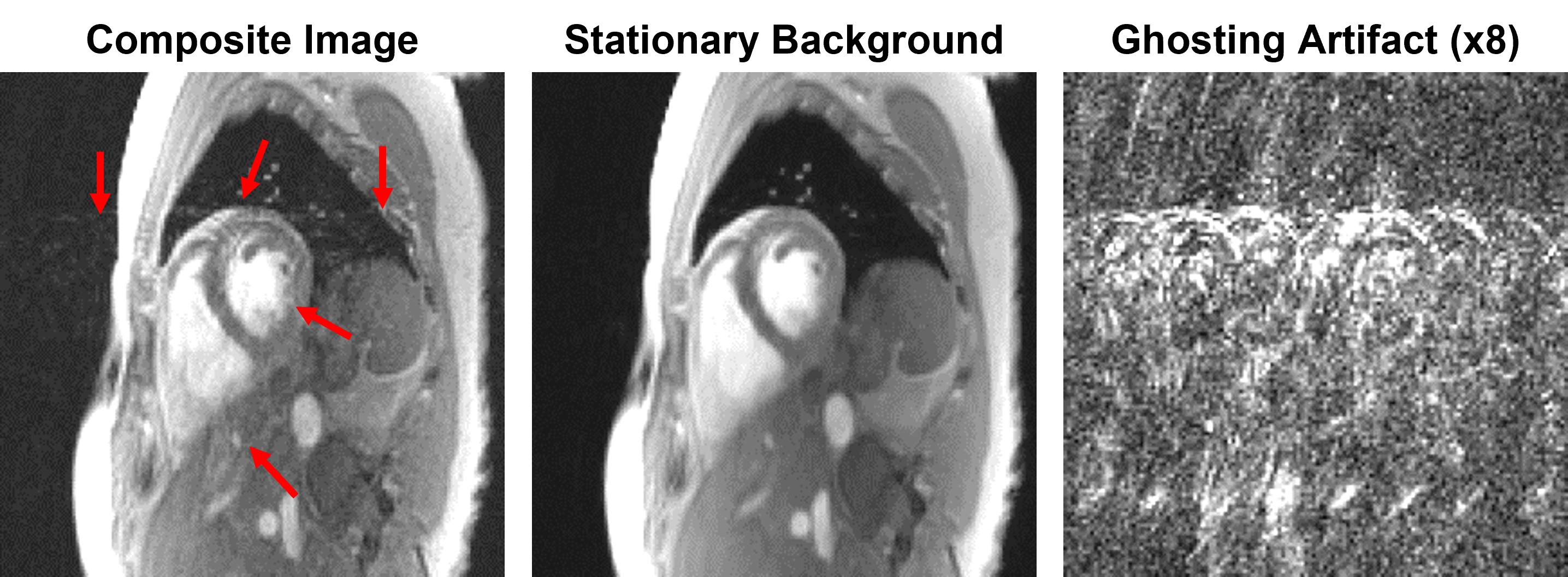}}
    \caption{Detection and removal of ghosting artifacts for the prospectively accelerated $R=8$ RT cine data using the proposed DL-based method. The composite image (left) contains both stationary and dynamic components, with motion-induced ghosting artifacts marked by red arrows. The stationary background (middle) is obtained by subtracting the estimated ghosting component, reducing aliasing and motion-related distortions. The ghosting artifact map (right) reveals the extracted ghosting patterns, exhibiting the expected pseudo-periodic structure. These results demonstrate the method’s effectiveness in detecting and isolating ghosting artifacts. Note there is no reference ghosting pattern in this case, since the dataset is prospectively accelerated to $R=8$.}
    \label{fig: ghosting_pros}
\end{figure}

Fig. \ref{fig: ghosting_pros} shows the ghosting detection performance for prospectively accelerated data. We note that there is no reference ghosting artifact available for prospectively $R=8$ accelerated data, thus we only depict the decomposition of the composite image to its estimated ghosting and stationary components using the trained ghost detection model. The composite image (left) contains both stationary and dynamic structures, with motion-related artifacts indicated by red arrows. The stationary background (middle) is obtained by subtracting the estimated ghosting component from the composite image, effectively reducing aliasing and motion-induced artifacts. The extracted ghosting artifacts (right) exhibit pseudo-periodic patterns, which align with expected cardiac motion-related ghosting. Although reference labels are unavailable in this prospectively accelerated study, the visual results demonstrate that the proposed approach successfully isolates structured ghosting artifacts, resulting in a cleaner background image.

\subsubsection{RT cine MRI reconstructions} 
\begin{figure*}[!t]
    \centerline{\includegraphics[width=\columnwidth]{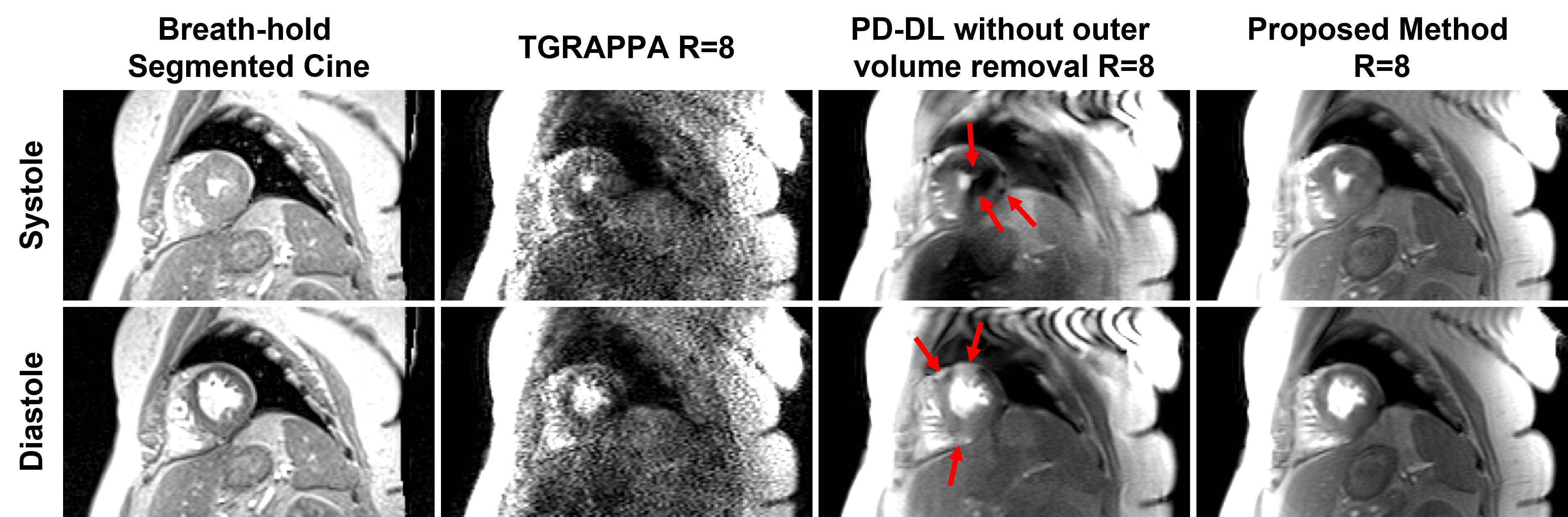}}
    \caption{Representative reconstruction results for the prospective $R=8$ RT cine dataset, comparing TGRAPPA at $R=8$, PD-DL without OVR, and the proposed method. The BH segmented cine acquisition from a different scan (left) is included as the clinical baseline. TGRAPPA at $R=8$ exhibits severe residual artifacts and low SNR. PD-DL without OVR reduces artifacts but introduces blurring and signal loss at the myocardium-blood interface (red arrows). The proposed method mitigates these issues, producing images with improved anatomical detail, comparable to the BH reference.}
    \label{pro_recon}
\end{figure*}

Fig.~\ref{pro_recon} shows the final reconstruction results for the prospectively accelerated $R$ = 8 RT cine data, reconstructed with TGRAPPA at $R$ = 8, PD-DL without OVR, and the proposed OVR-based method, as well as the BH segmented cine clinical baseline. Consistent with retrospective acceleration, TGRAPPA at $R$ = 8 exhibits severe residual artifacts and low signal-to-noise ratio (SNR), while PD-DL without OVR reduces artifacts but introduces blurring and signal loss at the myocardium-blood interface (red arrows). In contrast, the proposed method effectively mitigates these issues, producing images with improved anatomical detail, comparable to the BH reference, which was acquired in a separate scan as described in Sec. \ref{sec: prospective experiments}.

\subsubsection{Quantitative cardiac function analysis}
Table~\ref{tab2} summarizes the quantitative LV functional parameters, showing strong agreement between the proposed method and the BH segmented cine reference with no statistical differences ($P > .05$ in all cases). These results reinforce the effectiveness of the proposed method in highly-accelerated RT cine MRI.

\begin{table}[H]
\centering
\caption{\label{tab2} LV analysis for prospective study results. EDV: end diastolic volume; ESV: end systolic volume; EF: ejection fraction; SV: stroke volume; EDM: end diastolic mass; ESM: end systolic mass. Results are given as a mean, with the standard deviation in parentheses.}
\begin{tabular}{lll}
\hline
Metric & Proposed Method & BH Segmented Cine \\
\hline
EDV (mL)& 119.7 (17) & 121.3 (19) \\
ESV (mL)& 37.0 (11)  & 37.0 (13)  \\
EF (\%) & 69.7 (5)   & 70.2 (7)   \\
SV (mL) & 82.3 (8)   & 84.3 (8)   \\
EDM (g) & 138.0 (30) & 138.5 (34) \\
ESM (g) & 138.8 (25) & 139.0 (27) \\
\hline
\end{tabular}
\end{table}

\section{Discussion}
In this study, we introduced a novel outer volume removal and reconstruction strategy for highly-accelerated RT dynamic MRI. We first analytically characterized the pseudo-periodic ghosting artifacts that arise in low-temporal-resolution composite images formed from time-interleaved shifted undersampling patterns, demonstrating that these ghosting patterns originate from moving tissue, and propagate through time in a pseudo-periodic manner, creating structured interference within the field-of-view (FOV). Building upon this enhanced characterization, we developed a DL-based strategy to estimate and remove these motion-induced ghosting artifacts from composite images. Our DL model effectively differentiated structured ghosting artifacts from the true background, generating a clean estimate of the outer volume signal. This estimated signal was then subtracted from each individual timeframe’s k-space data before the final reconstruction, significantly reducing aliasing artifacts while maintaining temporal fidelity. We further proposed a novel loss function for self-supervised training of PD-DL reconstruction, leveraging insights on the coil sensitivities after OVR.  Our results demonstrated that our framework enables significantly higher acceleration rates while maintaining high image quality both with retrospective and prospective acceleration, outperforming a matched PD-DL method without OVR. The proposed strategy represents a novel and practical alternative to conventional outer volume suppression (OVS) methods that operate at the acquisition level. Unlike OVS, our OVR approach applies corrections in post-processing without requiring specialized pulse designs, offering flexibility for integration with existing real-time steady-state imaging protocols.

In this study, we focused on demonstrating the feasibility of the proposed approach through a two-step pipeline that first detects and subtracts the ghosting artifacts, and then performs PD-DL reconstruction. While this approach shows promising results, further performance improvements could be achieved by adopting an end-to-end training strategy to enable joint optimization of both ghosting detection and image reconstruction processes. End-to-end training may enhance the model's ability to learn more intricate relationships between artifacts and the underlying data, potentially leading to even better artifact reduction and improved image quality \citep{zhang2024direct, huang2019fr}.

For prospective acceleration, we opted for a GRE sequence at 3T. This was mainly because the reconstruction pipeline was not available on the scanner, thus shimming issues and off-resonance artifacts \citep{ferreira2013cardiovascular, haskell2023off} could not be visualized and remedied at acquisition time. By using GRE sequences, we aimed to mitigate these issues and ensure robust data acquisition. We note that bSSFP is considered the gold standard for cine imaging \citep{kramer2020standardized}, particularly due to its superior signal-to-noise ratio and image quality. Furthermore, the TR of bSSFP sequences is typically shorter, allowing for an improved temporal resolution. In spite of these differences, the results from our prospective study demonstrated reliable performance, with the proposed method showing strong agreement with the baseline BH reference, while also showing that our method generalizes well to both bSSFP and GRE-based steady-state sequences without issues.

Our evaluation in this study aims to characterize the benefits of our OVR strategies. As such, it does not focus on specific neural network architectures or algorithm unrolling designs. Thus, our strategies are, in principle, agnostic to the details of the PD-DL method. Similarly, all our comparisons focus on reconstructions with only spatial regularization without a temporal regularization component. This is by design, to ensure that the reconstruction itself does not introduce any temporal blurring/artifacts, so that we can pinpoint whether the OVR processing alone leads to such issues. As our study shows that the OVR processing does not incur temporal artifacts both visually and quantitatively, further studies to use OVR with spatiotemporally regularized reconstructions or DL methods \citep{van2018fetal, kofler2019spatio,  sandino2021accelerating,  terpstra2023accelerated, lin2023free, oscanoa2023deep, wang2024deep} are warranted. Finally, we note that our OVR technique offers complementary acceleration to ideas related to virtual coils \citep{kim2021region}, and may be combined with them for further gains, though this is not the focus of this study.

A key limitation in our evaluation and training is the absence of reference images for RT dynamic imaging, which is particularly important in cases that require high temporal resolution. Without reference data, our DL model for ghosting detection must rely on low-temporal-resolution images, which may not fully capture fast or subtle motion in certain anatomical regions. Though our prospective acceleration results show that ghosting detection trained on retrospectively accelerated $R$ = 4 data extends to prospectively accelerated $R$ = 8 data at twice improved temporal resolution, this limitation may affect accuracy in other applications, particularly when dealing with rapid or irregular tissue movements. As a result, the reduction in time resolution may lead to incomplete artifact removal or a loss of valuable details in highly dynamic structures and should be investigated for different applications.

While this study focuses on RT cine imaging, the proposed method has broader applicability to other dynamic imaging scenarios where the FOV extends significantly beyond the region of interest. Many imaging applications, such as abdominal dynamic MRI, functional MRI, and dynamic musculoskeletal imaging, suffer from similar challenges, where signals from surrounding tissues introduce aliasing artifacts and degrade image quality in the ROI \citep{smith2012reduced, kim2021region, tian2021aliasing}. By estimating and removing these outer volume signals in post-processing, our approach offers a framework for improving reconstruction quality in various accelerated MRI applications. \citet{tian2021aliasing} proposed a similar solution in spiral RT MRI for speech datasets by estimating and subtracting the source of aliasing from outer volume regions using large-FOV reconstruction and forward modeling. However, their method does not account for motion-induced ghosting artifacts in the outer volume, though we note that the effects of such artifacts in highly-accelerated spiral imaging need their own analytical characterization similar to our analysis here. Thus, future work to explore the adaptation of our method to different organs, motion patterns and sampling schemes is warranted.

\section{Conclusion}
In this study, we introduced a novel outer volume removal method for real-time dynamic MRI, aimed at reducing aliasing artifacts from extra-cardiac tissues without requiring modifications to the acquisition process. By leveraging composite temporal images formed from time-interleaved undersampling patterns, we demonstrated that pseudo-periodic ghosting artifacts can be accurately characterized and removed using a deep learning (DL) model, leading to a subsequent simpler reconstruction problem. Our results confirmed the superiority of our method over conventional and state-of-the-art techniques at $R=8$ acceleration both qualitatively and quantitatively. Importantly, since our method does not require specialized acquisition strategies, it can be readily integrated into existing clinical workflows. We will release our source codes upon acceptance of the paper.

\section*{Acknowledgment}
This work was partially supported by NIH R01HL153146, NIH R01EB032830, NIH P41EB027061, the European Research Council (101078711), the Nederlandse Hartstichting (03-004-2022-0079), and the Nederlandse Organisatie voor Wetenschappelijk Onderzoek (STU.019.024). The authors would like to thank Dr. Peter Kellman for providing the real-time cine MRI data and for insightful discussions. Additionally, the authors thank the Intramural Research Program of the National Heart, Lung, and Blood Institute for the data obtained from the NIH Open Source Cardiac MRI Raw Data Repository. A preliminary version of this work was partially presented at the 21st IEEE International Symposium on Biomedical Imaging (ISBI) \citep{gulle2024robust} and the 32nd Annual Meeting of the ISMRM \citep{Gulle2024}.

\bibliographystyle{johd}
\bibliography{references}

\begin{thebibliography}{}

\bibitem [\protect \citeauthoryear {%
Aggarwal%
\ \protect \BOthers {.}}{%
Aggarwal%
\ \protect \BOthers {.}}{%
{\protect \APACyear {2018}}%
}]{%
aggarwal2018modl}
\APACinsertmetastar {%
aggarwal2018modl}%
\begin{APACrefauthors}%
Aggarwal, H\BPBI K.%
, Mani, M\BPBI P.%
\BCBL {}\ \BBA {} Jacob, M.%
\end{APACrefauthors}%
\unskip\
\newblock
\APACrefYearMonthDay{2018}{}{}.
\newblock
{\BBOQ}\APACrefatitle {{MoDL}: Model-based deep learning architecture for inverse problems} {{MoDL}: Model-based deep learning architecture for inverse problems}.{\BBCQ}
\newblock
\APACjournalVolNumPages{IEEE Transactions on Medical Imaging}{38}{2}{394--405}.
\PrintBackRefs{\CurrentBib}

\bibitem [\protect \citeauthoryear {%
Ak{\c{c}}akaya%
\ \protect \BOthers {.}}{%
Ak{\c{c}}akaya%
\ \protect \BOthers {.}}{%
{\protect \APACyear {2022}}%
}]{%
akcakaya2022unsupervised}
\APACinsertmetastar {%
akcakaya2022unsupervised}%
\begin{APACrefauthors}%
Ak{\c{c}}akaya, M.%
, Yaman, B.%
, Chung, H.%
\BCBL {}\ \BBA {} Ye, J\BPBI C.%
\end{APACrefauthors}%
\unskip\
\newblock
\APACrefYearMonthDay{2022}{}{}.
\newblock
{\BBOQ}\APACrefatitle {Unsupervised deep learning methods for biological image reconstruction and enhancement} {Unsupervised deep learning methods for biological image reconstruction and enhancement}.{\BBCQ}
\newblock
\APACjournalVolNumPages{IEEE Signal Processing Magazine}{39}{2}{28--44}.
\PrintBackRefs{\CurrentBib}

\bibitem [\protect \citeauthoryear {%
Beer%
\ \protect \BOthers {.}}{%
Beer%
\ \protect \BOthers {.}}{%
{\protect \APACyear {2010}}%
}]{%
beer2010free}
\APACinsertmetastar {%
beer2010free}%
\begin{APACrefauthors}%
Beer, M.%
, Stamm, H.%
, Machann, W.%
, Weng, A.%
, Goltz, J\BPBI P.%
, Breunig, F.%
\BDBL {}K{\"o}stler, H.%
\end{APACrefauthors}%
\unskip\
\newblock
\APACrefYearMonthDay{2010}{}{}.
\newblock
{\BBOQ}\APACrefatitle {Free breathing cardiac real-time cine {MR} without {ECG} triggering} {Free breathing cardiac real-time cine {MR} without {ECG} triggering}.{\BBCQ}
\newblock
\APACjournalVolNumPages{International Journal of Cardiology}{145}{2}{380--382}.
\PrintBackRefs{\CurrentBib}

\bibitem [\protect \citeauthoryear {%
Blaimer%
\ \protect \BOthers {.}}{%
Blaimer%
\ \protect \BOthers {.}}{%
{\protect \APACyear {2008}}%
}]{%
blaimer2008sparse}
\APACinsertmetastar {%
blaimer2008sparse}%
\begin{APACrefauthors}%
Blaimer, M.%
, Breuer, F\BPBI A.%
, Jacob, P\BPBI M.%
, Kellman, P.%
\BCBL {}\ \BBA {} Griswold, M\BPBI A.%
\end{APACrefauthors}%
\unskip\
\newblock
\APACrefYearMonthDay{2008}{}{}.
\newblock
{\BBOQ}\APACrefatitle {A sparse {TSENSE} approach for improved dynamic parallel {MRI}} {A sparse {TSENSE} approach for improved dynamic parallel {MRI}}.{\BBCQ}
\newblock
\BIn{} \APACrefbtitle {Proceedings of the 17th International Society for Magnetic Resonance in Medicine ({ISMRM}).} {Proceedings of the 17th international society for magnetic resonance in medicine ({ISMRM}).}
\newblock
\APACrefnote{Abstract No. 1270}
\PrintBackRefs{\CurrentBib}

\bibitem [\protect \citeauthoryear {%
Breuer%
\ \protect \BOthers {.}}{%
Breuer%
\ \protect \BOthers {.}}{%
{\protect \APACyear {2005}}%
}]{%
breuer2005dynamic}
\APACinsertmetastar {%
breuer2005dynamic}%
\begin{APACrefauthors}%
Breuer, F\BPBI A.%
, Kellman, P.%
, Griswold, M\BPBI A.%
\BCBL {}\ \BBA {} Jakob, P\BPBI M.%
\end{APACrefauthors}%
\unskip\
\newblock
\APACrefYearMonthDay{2005}{}{}.
\newblock
{\BBOQ}\APACrefatitle {Dynamic autocalibrated parallel imaging using temporal {GRAPPA (TGRAPPA)}} {Dynamic autocalibrated parallel imaging using temporal {GRAPPA (TGRAPPA)}}.{\BBCQ}
\newblock
\APACjournalVolNumPages{Magnetic Resonance in Medicine}{53}{4}{981--985}.
\PrintBackRefs{\CurrentBib}

\bibitem [\protect \citeauthoryear {%
Chubb%
\ \protect \BOthers {.}}{%
Chubb%
\ \protect \BOthers {.}}{%
{\protect \APACyear {2017}}%
}]{%
chubb2017development}
\APACinsertmetastar {%
chubb2017development}%
\begin{APACrefauthors}%
Chubb, H.%
, Harrison, J\BPBI L.%
, Weiss, S.%
, Krueger, S.%
, Koken, P.%
, Bloch, L\BPBI {\O}.%
\BDBL {}Razavi, R\BPBI S.%
\end{APACrefauthors}%
\unskip\
\newblock
\APACrefYearMonthDay{2017}{}{}.
\newblock
{\BBOQ}\APACrefatitle {Development, preclinical validation, and clinical translation of a cardiac magnetic resonance-electrophysiology system with active catheter tracking for ablation of cardiac arrhythmia} {Development, preclinical validation, and clinical translation of a cardiac magnetic resonance-electrophysiology system with active catheter tracking for ablation of cardiac arrhythmia}.{\BBCQ}
\newblock
\APACjournalVolNumPages{JACC: Clinical Electrophysiology}{3}{2}{89--103}.
\PrintBackRefs{\CurrentBib}

\bibitem [\protect \citeauthoryear {%
Coristine%
\ \protect \BOthers {.}}{%
Coristine%
\ \protect \BOthers {.}}{%
{\protect \APACyear {2015}}%
}]{%
coristine2015combined}
\APACinsertmetastar {%
coristine2015combined}%
\begin{APACrefauthors}%
Coristine, A\BPBI J.%
, Van~Heeswijk, R\BPBI B.%
\BCBL {}\ \BBA {} Stuber, M.%
\end{APACrefauthors}%
\unskip\
\newblock
\APACrefYearMonthDay{2015}{}{}.
\newblock
{\BBOQ}\APACrefatitle {Combined {T}2-preparation and two-dimensional pencil-beam inner volume selection} {Combined {T}2-preparation and two-dimensional pencil-beam inner volume selection}.{\BBCQ}
\newblock
\APACjournalVolNumPages{Magnetic Resonance in Medicine}{74}{2}{529--536}.
\PrintBackRefs{\CurrentBib}

\bibitem [\protect \citeauthoryear {%
Demirel%
\ \protect \BOthers {.}}{%
Demirel%
\ \protect \BOthers {.}}{%
{\protect \APACyear {2021}}%
}]{%
demirel202120}
\APACinsertmetastar {%
demirel202120}%
\begin{APACrefauthors}%
Demirel, {\"O}\BPBI B.%
, Yaman, B.%
, Dowdle, L.%
, Moeller, S.%
, Vizioli, L.%
, Yacoub, E.%
\BDBL {}Ak{\c{c}}akaya, M.%
\end{APACrefauthors}%
\unskip\
\newblock
\APACrefYearMonthDay{2021}{}{}.
\newblock
{\BBOQ}\APACrefatitle {20-fold accelerated 7{T} f{MRI} using referenceless self-supervised deep learning reconstruction} {20-fold accelerated 7{T} f{MRI} using referenceless self-supervised deep learning reconstruction}.{\BBCQ}
\newblock
\BIn{} \APACrefbtitle {Proceedings of the 43rd Annual International Conference of the IEEE Engineering in Medicine \& Biology Society ({EMBC})} {Proceedings of the 43rd annual international conference of the ieee engineering in medicine \& biology society ({EMBC})}\ (\BPGS\ 3765--3769).
\PrintBackRefs{\CurrentBib}

\bibitem [\protect \citeauthoryear {%
Demirel%
, Yaman%
\BCBL {}\ \protect \BOthers {.}}{%
Demirel%
, Yaman%
\BCBL {}\ \protect \BOthers {.}}{%
{\protect \APACyear {2023}}%
}]{%
DemirelSIIM}
\APACinsertmetastar {%
DemirelSIIM}%
\begin{APACrefauthors}%
Demirel, {\"O}\BPBI B.%
, Yaman, B.%
, Shenoy, C.%
, Moeller, S.%
, Weingärtner, S.%
\BCBL {}\ \BBA {} Akçakaya, M.%
\end{APACrefauthors}%
\unskip\
\newblock
\APACrefYearMonthDay{2023}{}{}.
\newblock
{\BBOQ}\APACrefatitle {Signal intensity informed multi-coil encoding operator for physics-guided deep learning reconstruction of highly accelerated myocardial perfusion CMR} {Signal intensity informed multi-coil encoding operator for physics-guided deep learning reconstruction of highly accelerated myocardial perfusion cmr}.{\BBCQ}
\newblock
\APACjournalVolNumPages{Magnetic Resonance in Medicine}{89}{1}{308-321}.
\PrintBackRefs{\CurrentBib}

\bibitem [\protect \citeauthoryear {%
Demirel%
, Zhang%
\BCBL {}\ \protect \BOthers {.}}{%
Demirel%
, Zhang%
\BCBL {}\ \protect \BOthers {.}}{%
{\protect \APACyear {2023}}%
}]{%
demirel2023high}
\APACinsertmetastar {%
demirel2023high}%
\begin{APACrefauthors}%
Demirel, {\"O}\BPBI B.%
, Zhang, C.%
, Yaman, B.%
, G{\"u}lle, M.%
, Shenoy, C.%
, Leiner, T.%
\BDBL {}Ak{\c{c}}akaya, M.%
\end{APACrefauthors}%
\unskip\
\newblock
\APACrefYearMonthDay{2023}{}{}.
\newblock
{\BBOQ}\APACrefatitle {High-fidelity database-free deep learning reconstruction for real-time cine cardiac {MRI}} {High-fidelity database-free deep learning reconstruction for real-time cine cardiac {MRI}}.{\BBCQ}
\newblock
\BIn{} \APACrefbtitle {Proceedings of the 45th Annual International Conference of the IEEE Engineering in Medicine \& Biology Society (EMBC)} {Proceedings of the 45th annual international conference of the ieee engineering in medicine \& biology society (embc)}\ (\BPGS\ 1--4).
\PrintBackRefs{\CurrentBib}

\bibitem [\protect \citeauthoryear {%
Fan%
\ \protect \BOthers {.}}{%
Fan%
\ \protect \BOthers {.}}{%
{\protect \APACyear {2024}}%
}]{%
fan2024ultra}
\APACinsertmetastar {%
fan2024ultra}%
\begin{APACrefauthors}%
Fan, L.%
, Hong, K.%
, Allen, B\BPBI D.%
, Paul, R.%
, Carr, J\BPBI C.%
, Zhang, S.%
\BDBL {}others%
\end{APACrefauthors}%
\unskip\
\newblock
\APACrefYearMonthDay{2024}{}{}.
\newblock
{\BBOQ}\APACrefatitle {Ultra-rapid, Free-breathing, Real-time Cardiac Cine {MRI} Using {GRASP} Amplified with View Sharing and {KWIC} Filtering} {Ultra-rapid, free-breathing, real-time cardiac cine {MRI} using {GRASP} amplified with view sharing and {KWIC} filtering}.{\BBCQ}
\newblock
\APACjournalVolNumPages{Radiology: Cardiothoracic Imaging}{6}{1}{}.
\newblock
\APACrefnote{Art. no. e230107}
\PrintBackRefs{\CurrentBib}

\bibitem [\protect \citeauthoryear {%
Feng%
\ \protect \BOthers {.}}{%
Feng%
\ \protect \BOthers {.}}{%
{\protect \APACyear {2016}}%
}]{%
feng2016xd}
\APACinsertmetastar {%
feng2016xd}%
\begin{APACrefauthors}%
Feng, L.%
, Axel, L.%
, Chandarana, H.%
, Block, K\BPBI T.%
, Sodickson, D\BPBI K.%
\BCBL {}\ \BBA {} Otazo, R.%
\end{APACrefauthors}%
\unskip\
\newblock
\APACrefYearMonthDay{2016}{}{}.
\newblock
{\BBOQ}\APACrefatitle {{XD-GRASP}: golden-angle radial {MRI} with reconstruction of extra motion-state dimensions using compressed sensing} {{XD-GRASP}: golden-angle radial {MRI} with reconstruction of extra motion-state dimensions using compressed sensing}.{\BBCQ}
\newblock
\APACjournalVolNumPages{Magnetic Resonance in Medicine}{75}{2}{775--788}.
\PrintBackRefs{\CurrentBib}

\bibitem [\protect \citeauthoryear {%
Feng%
\ \protect \BOthers {.}}{%
Feng%
\ \protect \BOthers {.}}{%
{\protect \APACyear {2011}}%
}]{%
feng2011highly}
\APACinsertmetastar {%
feng2011highly}%
\begin{APACrefauthors}%
Feng, L.%
, Otazo, R.%
, Srichai, M\BPBI B.%
, Lim, R\BPBI P.%
, Sodickson, D\BPBI K.%
\BCBL {}\ \BBA {} Kim, D.%
\end{APACrefauthors}%
\unskip\
\newblock
\APACrefYearMonthDay{2011}{}{}.
\newblock
{\BBOQ}\APACrefatitle {Highly-accelerated real-time cine {MRI} using compressed sensing and parallel imaging} {Highly-accelerated real-time cine {MRI} using compressed sensing and parallel imaging}.{\BBCQ}
\newblock
\APACjournalVolNumPages{Journal of Cardiovascular Magnetic Resonance}{13}{}{1--2}.
\PrintBackRefs{\CurrentBib}

\bibitem [\protect \citeauthoryear {%
Ferreira%
\ \protect \BOthers {.}}{%
Ferreira%
\ \protect \BOthers {.}}{%
{\protect \APACyear {2013}}%
}]{%
ferreira2013cardiovascular}
\APACinsertmetastar {%
ferreira2013cardiovascular}%
\begin{APACrefauthors}%
Ferreira, P\BPBI F.%
, Gatehouse, P\BPBI D.%
, Mohiaddin, R\BPBI H.%
\BCBL {}\ \BBA {} Firmin, D\BPBI N.%
\end{APACrefauthors}%
\unskip\
\newblock
\APACrefYearMonthDay{2013}{}{}.
\newblock
{\BBOQ}\APACrefatitle {Cardiovascular magnetic resonance artefacts} {Cardiovascular magnetic resonance artefacts}.{\BBCQ}
\newblock
\APACjournalVolNumPages{Journal of Cardiovascular Magnetic Resonance}{15}{1}{41}.
\PrintBackRefs{\CurrentBib}

\bibitem [\protect \citeauthoryear {%
G{\'o}mez-Talavera%
\ \protect \BOthers {.}}{%
G{\'o}mez-Talavera%
\ \protect \BOthers {.}}{%
{\protect \APACyear {2021}}%
}]{%
gomez2021clinical}
\APACinsertmetastar {%
gomez2021clinical}%
\begin{APACrefauthors}%
G{\'o}mez-Talavera, S.%
, Fernandez-Jimenez, R.%
, Fuster, V.%
, Nothnagel, N\BPBI D.%
, Kouwenhoven, M.%
, Clemence, M.%
\BDBL {}others%
\end{APACrefauthors}%
\unskip\
\newblock
\APACrefYearMonthDay{2021}{}{}.
\newblock
{\BBOQ}\APACrefatitle {Clinical validation of a 3-dimensional ultrafast cardiac magnetic resonance protocol including single breath-hold 3-dimensional sequences} {Clinical validation of a 3-dimensional ultrafast cardiac magnetic resonance protocol including single breath-hold 3-dimensional sequences}.{\BBCQ}
\newblock
\APACjournalVolNumPages{JACC Cardiovasc Imaging}{14}{9}{1742--1754}.
\PrintBackRefs{\CurrentBib}

\bibitem [\protect \citeauthoryear {%
Griswold%
\ \protect \BOthers {.}}{%
Griswold%
\ \protect \BOthers {.}}{%
{\protect \APACyear {2004}}%
}]{%
griswold2004field}
\APACinsertmetastar {%
griswold2004field}%
\begin{APACrefauthors}%
Griswold, M\BPBI A.%
, Kannengiesser, S.%
, Heidemann, R\BPBI M.%
, Wang, J.%
\BCBL {}\ \BBA {} Jakob, P\BPBI M.%
\end{APACrefauthors}%
\unskip\
\newblock
\APACrefYearMonthDay{2004}{}{}.
\newblock
{\BBOQ}\APACrefatitle {Field-of-view limitations in parallel imaging} {Field-of-view limitations in parallel imaging}.{\BBCQ}
\newblock
\APACjournalVolNumPages{Magnetic Resonance in Medicine}{52}{5}{1118--1126}.
\PrintBackRefs{\CurrentBib}

\bibitem [\protect \citeauthoryear {%
Gu%
\ \protect \BOthers {.}}{%
Gu%
\ \protect \BOthers {.}}{%
{\protect \APACyear {2022}}%
}]{%
gu2022revisiting}
\APACinsertmetastar {%
gu2022revisiting}%
\begin{APACrefauthors}%
Gu, H.%
, Yaman, B.%
, Moeller, S.%
, Ellermann, J.%
, U{\u{g}}urbil, K.%
\BCBL {}\ \BBA {} Ak{\c{c}}akaya, M.%
\end{APACrefauthors}%
\unskip\
\newblock
\APACrefYearMonthDay{2022}{}{}.
\newblock
{\BBOQ}\APACrefatitle {Revisiting l1-wavelet compressed-sensing {MRI} in the era of deep learning} {Revisiting l1-wavelet compressed-sensing {MRI} in the era of deep learning}.{\BBCQ}
\newblock
\APACjournalVolNumPages{Proceedings of the National Academy of Sciences}{119}{33}{}.
\newblock
\APACrefnote{Art. no. e2201062119}
\PrintBackRefs{\CurrentBib}

\bibitem [\protect \citeauthoryear {%
G{\"u}lle%
\ \BBA {} Ak{\c{c}}akaya%
}{%
G{\"u}lle%
\ \BBA {} Ak{\c{c}}akaya%
}{%
{\protect \APACyear {2024}}%
}]{%
gulle2024robust}
\APACinsertmetastar {%
gulle2024robust}%
\begin{APACrefauthors}%
G{\"u}lle, M.%
\BCBT {}\ \BBA {} Ak{\c{c}}akaya, M.%
\end{APACrefauthors}%
\unskip\
\newblock
\APACrefYearMonthDay{2024}{}{}.
\newblock
{\BBOQ}\APACrefatitle {Robust Outer Volume Subtraction with Deep Learning Ghosting Detection for Highly-Accelerated Real-Time Dynamic {MRI}} {Robust outer volume subtraction with deep learning ghosting detection for highly-accelerated real-time dynamic {MRI}}.{\BBCQ}
\newblock
\BIn{} \APACrefbtitle {Proceedings of the 21st IEEE International Symposium on Biomedical Imaging ({ISBI})} {Proceedings of the 21st ieee international symposium on biomedical imaging ({ISBI})}\ (\BPGS\ 1--5).
\PrintBackRefs{\CurrentBib}

\bibitem [\protect \citeauthoryear {%
G{\"u}lle%
\ \protect \BOthers {.}}{%
G{\"u}lle%
\ \protect \BOthers {.}}{%
{\protect \APACyear {2024}}%
}]{%
Gulle2024}
\APACinsertmetastar {%
Gulle2024}%
\begin{APACrefauthors}%
G{\"u}lle, M.%
, Kellman, P.%
\BCBL {}\ \BBA {} Ak{\c{c}}akaya, M.%
\end{APACrefauthors}%
\unskip\
\newblock
\APACrefYearMonthDay{2024}{}{}.
\newblock
{\BBOQ}\APACrefatitle {Revisiting outer volume subtraction with deep-learning tools for highly-accelerated real-time cine {CMR}} {Revisiting outer volume subtraction with deep-learning tools for highly-accelerated real-time cine {CMR}}.{\BBCQ}
\newblock
\BIn{} \APACrefbtitle {Proceedings of the 33rd International Society for Magnetic Resonance in Medicine ({ISMRM}).} {Proceedings of the 33rd international society for magnetic resonance in medicine ({ISMRM}).}
\newblock
\begin{APACrefURL} \url{https://archive.ismrm.org/2024/1371.html} \end{APACrefURL}
\newblock
\APACrefnote{Abstract No. 1371}
\PrintBackRefs{\CurrentBib}

\bibitem [\protect \citeauthoryear {%
Hammernik%
\ \protect \BOthers {.}}{%
Hammernik%
\ \protect \BOthers {.}}{%
{\protect \APACyear {2018}}%
}]{%
hammernik2018learning}
\APACinsertmetastar {%
hammernik2018learning}%
\begin{APACrefauthors}%
Hammernik, K.%
, Klatzer, T.%
, Kobler, E.%
, Recht, M\BPBI P.%
, Sodickson, D\BPBI K.%
, Pock, T.%
\BCBL {}\ \BBA {} Knoll, F.%
\end{APACrefauthors}%
\unskip\
\newblock
\APACrefYearMonthDay{2018}{}{}.
\newblock
{\BBOQ}\APACrefatitle {Learning a variational network for reconstruction of accelerated {MRI} data} {Learning a variational network for reconstruction of accelerated {MRI} data}.{\BBCQ}
\newblock
\APACjournalVolNumPages{Magnetic Resonance in Medicine}{79}{6}{3055--3071}.
\PrintBackRefs{\CurrentBib}

\bibitem [\protect \citeauthoryear {%
Hammernik%
\ \protect \BOthers {.}}{%
Hammernik%
\ \protect \BOthers {.}}{%
{\protect \APACyear {2023}}%
}]{%
hammernik2023physics}
\APACinsertmetastar {%
hammernik2023physics}%
\begin{APACrefauthors}%
Hammernik, K.%
, K{\"u}stner, T.%
, Yaman, B.%
, Huang, Z.%
, Rueckert, D.%
, Knoll, F.%
\BCBL {}\ \BBA {} Ak{\c{c}}akaya, M.%
\end{APACrefauthors}%
\unskip\
\newblock
\APACrefYearMonthDay{2023}{Jan}{}.
\newblock
{\BBOQ}\APACrefatitle {{{P}hysics-driven deep learning for computational magnetic resonance imaging: combining physics and machine learning for improved medical imaging}} {{{P}hysics-driven deep learning for computational magnetic resonance imaging: combining physics and machine learning for improved medical imaging}}.{\BBCQ}
\newblock
\APACjournalVolNumPages{IEEE Signal Processing Magazine}{40}{1}{98--114}.
\PrintBackRefs{\CurrentBib}

\bibitem [\protect \citeauthoryear {%
Hansen%
\ \protect \BOthers {.}}{%
Hansen%
\ \protect \BOthers {.}}{%
{\protect \APACyear {2012}}%
}]{%
hansen2012retrospective}
\APACinsertmetastar {%
hansen2012retrospective}%
\begin{APACrefauthors}%
Hansen, M\BPBI S.%
, S{\o}rensen, T\BPBI S.%
, Arai, A\BPBI E.%
\BCBL {}\ \BBA {} Kellman, P.%
\end{APACrefauthors}%
\unskip\
\newblock
\APACrefYearMonthDay{2012}{}{}.
\newblock
{\BBOQ}\APACrefatitle {Retrospective reconstruction of high temporal resolution cine images from real-time {MRI} using iterative motion correction} {Retrospective reconstruction of high temporal resolution cine images from real-time {MRI} using iterative motion correction}.{\BBCQ}
\newblock
\APACjournalVolNumPages{Magnetic Resonance in Medicine}{68}{3}{741--750}.
\PrintBackRefs{\CurrentBib}

\bibitem [\protect \citeauthoryear {%
Haskell%
\ \protect \BOthers {.}}{%
Haskell%
\ \protect \BOthers {.}}{%
{\protect \APACyear {2023}}%
}]{%
haskell2023off}
\APACinsertmetastar {%
haskell2023off}%
\begin{APACrefauthors}%
Haskell, M\BPBI W.%
, Nielsen, J\BHBI F.%
\BCBL {}\ \BBA {} Noll, D\BPBI C.%
\end{APACrefauthors}%
\unskip\
\newblock
\APACrefYearMonthDay{2023}{}{}.
\newblock
{\BBOQ}\APACrefatitle {Off-resonance artifact correction for {MRI}: A review} {Off-resonance artifact correction for {MRI}: A review}.{\BBCQ}
\newblock
\APACjournalVolNumPages{Nuclear Magnetic Resonance in Biomedicine}{36}{5}{}.
\newblock
\APACrefnote{Art. no. e4867}
\PrintBackRefs{\CurrentBib}

\bibitem [\protect \citeauthoryear {%
Hauptmann%
\ \protect \BOthers {.}}{%
Hauptmann%
\ \protect \BOthers {.}}{%
{\protect \APACyear {2019}}%
}]{%
hauptmann2019real}
\APACinsertmetastar {%
hauptmann2019real}%
\begin{APACrefauthors}%
Hauptmann, A.%
, Arridge, S.%
, Lucka, F.%
, Muthurangu, V.%
\BCBL {}\ \BBA {} Steeden, J\BPBI A.%
\end{APACrefauthors}%
\unskip\
\newblock
\APACrefYearMonthDay{2019}{}{}.
\newblock
{\BBOQ}\APACrefatitle {Real-time cardiovascular {MR} with spatio-temporal artifact suppression using deep learning--proof of concept in congenital heart disease} {Real-time cardiovascular {MR} with spatio-temporal artifact suppression using deep learning--proof of concept in congenital heart disease}.{\BBCQ}
\newblock
\APACjournalVolNumPages{Magnetic Resonance in Medicine}{81}{2}{1143--1156}.
\PrintBackRefs{\CurrentBib}

\bibitem [\protect \citeauthoryear {%
Heiberg%
\ \protect \BOthers {.}}{%
Heiberg%
\ \protect \BOthers {.}}{%
{\protect \APACyear {2010}}%
}]{%
heiberg2010design}
\APACinsertmetastar {%
heiberg2010design}%
\begin{APACrefauthors}%
Heiberg, E.%
, Sj{\"o}gren, J.%
, Ugander, M.%
, Carlsson, M.%
, Engblom, H.%
\BCBL {}\ \BBA {} Arheden, H.%
\end{APACrefauthors}%
\unskip\
\newblock
\APACrefYearMonthDay{2010}{}{}.
\newblock
{\BBOQ}\APACrefatitle {Design and validation of segment-freely available software for cardiovascular image analysis} {Design and validation of segment-freely available software for cardiovascular image analysis}.{\BBCQ}
\newblock
\APACjournalVolNumPages{BMC Medical Imaging}{10}{}{1--13}.
\PrintBackRefs{\CurrentBib}

\bibitem [\protect \citeauthoryear {%
Hosseini%
\ \protect \BOthers {.}}{%
Hosseini%
\ \protect \BOthers {.}}{%
{\protect \APACyear {2020}}%
}]{%
hosseini2020dense}
\APACinsertmetastar {%
hosseini2020dense}%
\begin{APACrefauthors}%
Hosseini, S\BPBI A\BPBI H.%
, Yaman, B.%
, Moeller, S.%
, Hong, M.%
\BCBL {}\ \BBA {} Ak{\c{c}}akaya, M.%
\end{APACrefauthors}%
\unskip\
\newblock
\APACrefYearMonthDay{2020}{}{}.
\newblock
{\BBOQ}\APACrefatitle {Dense recurrent neural networks for accelerated {MRI}: History-cognizant unrolling of optimization algorithms} {Dense recurrent neural networks for accelerated {MRI}: History-cognizant unrolling of optimization algorithms}.{\BBCQ}
\newblock
\APACjournalVolNumPages{IEEE Journal of Selected Topics in Signal Processing}{14}{6}{1280--1291}.
\PrintBackRefs{\CurrentBib}

\bibitem [\protect \citeauthoryear {%
Huang%
\ \protect \BOthers {.}}{%
Huang%
\ \protect \BOthers {.}}{%
{\protect \APACyear {2019}}%
}]{%
huang2019fr}
\APACinsertmetastar {%
huang2019fr}%
\begin{APACrefauthors}%
Huang, Q.%
, Yang, D.%
, Yi, J.%
, Axel, L.%
\BCBL {}\ \BBA {} Metaxas, D.%
\end{APACrefauthors}%
\unskip\
\newblock
\APACrefYearMonthDay{2019}{}{}.
\newblock
{\BBOQ}\APACrefatitle {{FR-Net}: joint reconstruction and segmentation in compressed sensing cardiac {MRI}} {{FR-Net}: joint reconstruction and segmentation in compressed sensing cardiac {MRI}}.{\BBCQ}
\newblock
\BIn{} \APACrefbtitle {Proceedings of the 10th International Conference of the Functional Imaging and Modeling of the Heart ({FIMH})} {Proceedings of the 10th international conference of the functional imaging and modeling of the heart ({FIMH})}\ (\BPGS\ 352--360).
\PrintBackRefs{\CurrentBib}

\bibitem [\protect \citeauthoryear {%
Ishida%
\ \protect \BOthers {.}}{%
Ishida%
\ \protect \BOthers {.}}{%
{\protect \APACyear {2009}}%
}]{%
ishida2009cardiac}
\APACinsertmetastar {%
ishida2009cardiac}%
\begin{APACrefauthors}%
Ishida, M.%
, Kato, S.%
\BCBL {}\ \BBA {} Sakuma, H.%
\end{APACrefauthors}%
\unskip\
\newblock
\APACrefYearMonthDay{2009}{}{}.
\newblock
{\BBOQ}\APACrefatitle {Cardiac {MRI} in ischemic heart disease} {Cardiac {MRI} in ischemic heart disease}.{\BBCQ}
\newblock
\APACjournalVolNumPages{Circulation Journal}{73}{9}{1577--1588}.
\PrintBackRefs{\CurrentBib}

\bibitem [\protect \citeauthoryear {%
Jung%
\ \protect \BOthers {.}}{%
Jung%
\ \protect \BOthers {.}}{%
{\protect \APACyear {2009}}%
}]{%
jung2009k}
\APACinsertmetastar {%
jung2009k}%
\begin{APACrefauthors}%
Jung, H.%
, Sung, K.%
, Nayak, K\BPBI S.%
, Kim, E\BPBI Y.%
\BCBL {}\ \BBA {} Ye, J\BPBI C.%
\end{APACrefauthors}%
\unskip\
\newblock
\APACrefYearMonthDay{2009}{}{}.
\newblock
{\BBOQ}\APACrefatitle {k-t {FOCUSS}: a general compressed sensing framework for high resolution dynamic {MRI}} {k-t {FOCUSS}: a general compressed sensing framework for high resolution dynamic {MRI}}.{\BBCQ}
\newblock
\APACjournalVolNumPages{Magnetic Resonance in Medicine}{61}{1}{103--116}.
\PrintBackRefs{\CurrentBib}

\bibitem [\protect \citeauthoryear {%
Kawchuk%
\ \protect \BOthers {.}}{%
Kawchuk%
\ \protect \BOthers {.}}{%
{\protect \APACyear {2015}}%
}]{%
Kawchuk2015}
\APACinsertmetastar {%
Kawchuk2015}%
\begin{APACrefauthors}%
Kawchuk, G\BPBI N.%
, Fryer, J.%
, Jaremko, J\BPBI L.%
, Zeng, H.%
, Rowe, L.%
\BCBL {}\ \BBA {} Thompson, R.%
\end{APACrefauthors}%
\unskip\
\newblock
\APACrefYearMonthDay{2015}{}{}.
\newblock
{\BBOQ}\APACrefatitle {{{R}eal-time visualization of joint cavitation}} {{{R}eal-time visualization of joint cavitation}}.{\BBCQ}
\newblock
\APACjournalVolNumPages{PLoS One}{10}{4}{}.
\newblock
\APACrefnote{Art. no. e0119470}
\PrintBackRefs{\CurrentBib}

\bibitem [\protect \citeauthoryear {%
Kellman%
\ \protect \BOthers {.}}{%
Kellman%
\ \protect \BOthers {.}}{%
{\protect \APACyear {2001}}%
}]{%
kellman2001adaptive}
\APACinsertmetastar {%
kellman2001adaptive}%
\begin{APACrefauthors}%
Kellman, P.%
, Epstein, F\BPBI H.%
\BCBL {}\ \BBA {} McVeigh, E\BPBI R.%
\end{APACrefauthors}%
\unskip\
\newblock
\APACrefYearMonthDay{2001}{}{}.
\newblock
{\BBOQ}\APACrefatitle {Adaptive sensitivity encoding incorporating temporal filtering ({TSENSE})} {Adaptive sensitivity encoding incorporating temporal filtering ({TSENSE})}.{\BBCQ}
\newblock
\APACjournalVolNumPages{Magnetic Resonance in Medicine}{45}{5}{846--852}.
\PrintBackRefs{\CurrentBib}

\bibitem [\protect \citeauthoryear {%
Kim%
\ \protect \BOthers {.}}{%
Kim%
\ \protect \BOthers {.}}{%
{\protect \APACyear {2021}}%
}]{%
kim2021region}
\APACinsertmetastar {%
kim2021region}%
\begin{APACrefauthors}%
Kim, D.%
, Cauley, S\BPBI F.%
, Nayak, K\BPBI S.%
, Leahy, R\BPBI M.%
\BCBL {}\ \BBA {} Haldar, J\BPBI P.%
\end{APACrefauthors}%
\unskip\
\newblock
\APACrefYearMonthDay{2021}{}{}.
\newblock
{\BBOQ}\APACrefatitle {Region-optimized virtual {(ROVir)} coils: Localization and/or suppression of spatial regions using sensor-domain beamforming} {Region-optimized virtual {(ROVir)} coils: Localization and/or suppression of spatial regions using sensor-domain beamforming}.{\BBCQ}
\newblock
\APACjournalVolNumPages{Magnetic Resonance in Medicine}{86}{1}{197--212}.
\PrintBackRefs{\CurrentBib}

\bibitem [\protect \citeauthoryear {%
Knoll%
\ \protect \BOthers {.}}{%
Knoll%
\ \protect \BOthers {.}}{%
{\protect \APACyear {2020}}%
}]{%
Knoll_SPM}
\APACinsertmetastar {%
Knoll_SPM}%
\begin{APACrefauthors}%
Knoll, F.%
, Hammernik, K.%
, Zhang, C.%
, Moeller, S.%
, Pock, T.%
, Sodickson, D\BPBI K.%
\BCBL {}\ \BBA {} Ak{\c{c}}akaya, M.%
\end{APACrefauthors}%
\unskip\
\newblock
\APACrefYearMonthDay{2020}{}{}.
\newblock
{\BBOQ}\APACrefatitle {{{D}eep learning methods for parallel magnetic resonance imaging reconstruction}} {{{D}eep learning methods for parallel magnetic resonance imaging reconstruction}}.{\BBCQ}
\newblock
\APACjournalVolNumPages{IEEE Signal Processing Magazine}{37}{}{128--140}.
\PrintBackRefs{\CurrentBib}

\bibitem [\protect \citeauthoryear {%
Kofler%
\ \protect \BOthers {.}}{%
Kofler%
\ \protect \BOthers {.}}{%
{\protect \APACyear {2019}}%
}]{%
kofler2019spatio}
\APACinsertmetastar {%
kofler2019spatio}%
\begin{APACrefauthors}%
Kofler, A.%
, Dewey, M.%
, Schaeffter, T.%
, Wald, C.%
\BCBL {}\ \BBA {} Kolbitsch, C.%
\end{APACrefauthors}%
\unskip\
\newblock
\APACrefYearMonthDay{2019}{}{}.
\newblock
{\BBOQ}\APACrefatitle {Spatio-temporal deep learning-based undersampling artefact reduction for 2{D} radial cine {MRI} with limited training data} {Spatio-temporal deep learning-based undersampling artefact reduction for 2{D} radial cine {MRI} with limited training data}.{\BBCQ}
\newblock
\APACjournalVolNumPages{IEEE Transactions on Medical Imaging}{39}{3}{703--717}.
\PrintBackRefs{\CurrentBib}

\bibitem [\protect \citeauthoryear {%
Kramer%
\ \protect \BOthers {.}}{%
Kramer%
\ \protect \BOthers {.}}{%
{\protect \APACyear {2020}}%
}]{%
kramer2020standardized}
\APACinsertmetastar {%
kramer2020standardized}%
\begin{APACrefauthors}%
Kramer, C\BPBI M.%
, Barkhausen, J.%
, Bucciarelli-Ducci, C.%
, Flamm, S\BPBI D.%
, Kim, R\BPBI J.%
\BCBL {}\ \BBA {} Nagel, E.%
\end{APACrefauthors}%
\unskip\
\newblock
\APACrefYearMonthDay{2020}{}{}.
\newblock
{\BBOQ}\APACrefatitle {Standardized cardiovascular magnetic resonance imaging ({CMR}) protocols: 2020 update} {Standardized cardiovascular magnetic resonance imaging ({CMR}) protocols: 2020 update}.{\BBCQ}
\newblock
\APACjournalVolNumPages{Journal of Cardiovascular Magnetic Resonance}{22}{1}{17}.
\PrintBackRefs{\CurrentBib}

\bibitem [\protect \citeauthoryear {%
Li%
\ \protect \BOthers {.}}{%
Li%
\ \protect \BOthers {.}}{%
{\protect \APACyear {2018}}%
}]{%
li2018real}
\APACinsertmetastar {%
li2018real}%
\begin{APACrefauthors}%
Li, Y\BPBI Y.%
, Rashid, S.%
, Cheng, Y\BPBI J.%
, Schapiro, W.%
, Gliganic, K.%
, Yamashita, A\BHBI M.%
\BDBL {}others%
\end{APACrefauthors}%
\unskip\
\newblock
\APACrefYearMonthDay{2018}{}{}.
\newblock
{\BBOQ}\APACrefatitle {Real-time cardiac {MRI} with radial acquisition and k-space variant reduced-{FOV} reconstruction} {Real-time cardiac {MRI} with radial acquisition and k-space variant reduced-{FOV} reconstruction}.{\BBCQ}
\newblock
\APACjournalVolNumPages{Magnetic Resonance Imaging}{53}{}{98--104}.
\PrintBackRefs{\CurrentBib}

\bibitem [\protect \citeauthoryear {%
Lin%
\ \protect \BOthers {.}}{%
Lin%
\ \protect \BOthers {.}}{%
{\protect \APACyear {2023}}%
}]{%
lin2023free}
\APACinsertmetastar {%
lin2023free}%
\begin{APACrefauthors}%
Lin, L.%
, Li, Y.%
, Wang, J.%
, Cao, L.%
, Liu, Y.%
, Pang, J.%
\BDBL {}Wang, Y.%
\end{APACrefauthors}%
\unskip\
\newblock
\APACrefYearMonthDay{2023}{}{}.
\newblock
{\BBOQ}\APACrefatitle {Free-breathing cardiac cine {MRI} with compressed sensing real-time imaging and retrospective motion correction: clinical feasibility and validation} {Free-breathing cardiac cine {MRI} with compressed sensing real-time imaging and retrospective motion correction: clinical feasibility and validation}.{\BBCQ}
\newblock
\APACjournalVolNumPages{European Radiology}{33}{4}{2289--2300}.
\PrintBackRefs{\CurrentBib}

\bibitem [\protect \citeauthoryear {%
Lingala%
\ \protect \BOthers {.}}{%
Lingala%
\ \protect \BOthers {.}}{%
{\protect \APACyear {2016}}%
}]{%
Lingala2016}
\APACinsertmetastar {%
Lingala2016}%
\begin{APACrefauthors}%
Lingala, S\BPBI G.%
, Sutton, B\BPBI P.%
, Miquel, M\BPBI E.%
\BCBL {}\ \BBA {} Nayak, K\BPBI S.%
\end{APACrefauthors}%
\unskip\
\newblock
\APACrefYearMonthDay{2016}{Jan}{}.
\newblock
{\BBOQ}\APACrefatitle {{{R}ecommendations for real-time speech {M}{R}{I}}} {{{R}ecommendations for real-time speech {M}{R}{I}}}.{\BBCQ}
\newblock
\APACjournalVolNumPages{Journal of Magnetic Resonance Imaging}{43}{1}{28--44}.
\PrintBackRefs{\CurrentBib}

\bibitem [\protect \citeauthoryear {%
Liu%
\ \protect \BOthers {.}}{%
Liu%
\ \protect \BOthers {.}}{%
{\protect \APACyear {2020}}%
}]{%
liu2020dynamic}
\APACinsertmetastar {%
liu2020dynamic}%
\begin{APACrefauthors}%
Liu, F.%
, Li, D.%
, Jin, X.%
, Qiu, W.%
, Xia, Q.%
\BCBL {}\ \BBA {} Sun, B.%
\end{APACrefauthors}%
\unskip\
\newblock
\APACrefYearMonthDay{2020}{}{}.
\newblock
{\BBOQ}\APACrefatitle {Dynamic cardiac {MRI} reconstruction using motion aligned locally low rank tensor ({MALLRT})} {Dynamic cardiac {MRI} reconstruction using motion aligned locally low rank tensor ({MALLRT})}.{\BBCQ}
\newblock
\APACjournalVolNumPages{Magnetic Resonance Imaging}{66}{}{104--115}.
\PrintBackRefs{\CurrentBib}

\bibitem [\protect \citeauthoryear {%
Long{\`e}re%
\ \protect \BOthers {.}}{%
Long{\`e}re%
\ \protect \BOthers {.}}{%
{\protect \APACyear {2023}}%
}]{%
longere2023new}
\APACinsertmetastar {%
longere2023new}%
\begin{APACrefauthors}%
Long{\`e}re, B.%
, Abassebay, N.%
, Gkizas, C.%
, Hennicaux, J.%
, Simeone, A.%
, Musso, A\BPBI R.%
\BDBL {}others%
\end{APACrefauthors}%
\unskip\
\newblock
\APACrefYearMonthDay{2023}{}{}.
\newblock
{\BBOQ}\APACrefatitle {A new compressed sensing cine cardiac MRI sequence with free-breathing real-time acquisition and fully automated motion-correction: a comprehensive evaluation} {A new compressed sensing cine cardiac mri sequence with free-breathing real-time acquisition and fully automated motion-correction: a comprehensive evaluation}.{\BBCQ}
\newblock
\APACjournalVolNumPages{Diagnostic and Interventional Imaging}{104}{11}{538--546}.
\PrintBackRefs{\CurrentBib}

\bibitem [\protect \citeauthoryear {%
Luo%
\ \protect \BOthers {.}}{%
Luo%
\ \protect \BOthers {.}}{%
{\protect \APACyear {2015}}%
}]{%
luo2015combined}
\APACinsertmetastar {%
luo2015combined}%
\begin{APACrefauthors}%
Luo, J.%
, Addy, N\BPBI O.%
, Ingle, R\BPBI R.%
, Hargreaves, B\BPBI A.%
, Hu, B\BPBI S.%
, Nishimura, D\BPBI G.%
\BCBL {}\ \BBA {} Shin, T.%
\end{APACrefauthors}%
\unskip\
\newblock
\APACrefYearMonthDay{2015}{}{}.
\newblock
{\BBOQ}\APACrefatitle {Combined outer volume suppression and {T}2 preparation sequence for coronary angiography} {Combined outer volume suppression and {T}2 preparation sequence for coronary angiography}.{\BBCQ}
\newblock
\APACjournalVolNumPages{Magnetic Resonance in Medicine}{74}{6}{1632--1639}.
\PrintBackRefs{\CurrentBib}

\bibitem [\protect \citeauthoryear {%
Najeeb%
\ \protect \BOthers {.}}{%
Najeeb%
\ \protect \BOthers {.}}{%
{\protect \APACyear {2020}}%
}]{%
najeeb2020respiratory}
\APACinsertmetastar {%
najeeb2020respiratory}%
\begin{APACrefauthors}%
Najeeb, F.%
, Usman, M.%
, Aslam, I.%
, Qazi, S\BPBI A.%
\BCBL {}\ \BBA {} Omer, H.%
\end{APACrefauthors}%
\unskip\
\newblock
\APACrefYearMonthDay{2020}{}{}.
\newblock
{\BBOQ}\APACrefatitle {Respiratory motion-corrected, compressively sampled dynamic MR image reconstruction by exploiting multiple sparsity constraints and phase correlation-based data binning} {Respiratory motion-corrected, compressively sampled dynamic mr image reconstruction by exploiting multiple sparsity constraints and phase correlation-based data binning}.{\BBCQ}
\newblock
\APACjournalVolNumPages{Magnetic Resonance Materials in Physics, Biology and Medicine}{33}{}{411--419}.
\PrintBackRefs{\CurrentBib}

\bibitem [\protect \citeauthoryear {%
Nayak%
\ \protect \BOthers {.}}{%
Nayak%
\ \protect \BOthers {.}}{%
{\protect \APACyear {2022}}%
}]{%
nayak2022}
\APACinsertmetastar {%
nayak2022}%
\begin{APACrefauthors}%
Nayak, K\BPBI S.%
, Lim, Y.%
, Campbell-Washburn, A\BPBI E.%
\BCBL {}\ \BBA {} Steeden, J.%
\end{APACrefauthors}%
\unskip\
\newblock
\APACrefYearMonthDay{2022}{}{}.
\newblock
{\BBOQ}\APACrefatitle {Real-time magnetic resonance imaging} {Real-time magnetic resonance imaging}.{\BBCQ}
\newblock
\APACjournalVolNumPages{Journal of Magnetic Resonance Imaging}{55}{1}{81--99}.
\PrintBackRefs{\CurrentBib}

\bibitem [\protect \citeauthoryear {%
Oscanoa%
\ \protect \BOthers {.}}{%
Oscanoa%
\ \protect \BOthers {.}}{%
{\protect \APACyear {2023}}%
}]{%
oscanoa2023deep}
\APACinsertmetastar {%
oscanoa2023deep}%
\begin{APACrefauthors}%
Oscanoa, J\BPBI A.%
, Middione, M\BPBI J.%
, Alkan, C.%
, Yurt, M.%
, Loecher, M.%
, Vasanawala, S\BPBI S.%
\BCBL {}\ \BBA {} Ennis, D\BPBI B.%
\end{APACrefauthors}%
\unskip\
\newblock
\APACrefYearMonthDay{2023}{}{}.
\newblock
{\BBOQ}\APACrefatitle {Deep learning-based reconstruction for cardiac MRI: a review} {Deep learning-based reconstruction for cardiac mri: a review}.{\BBCQ}
\newblock
\APACjournalVolNumPages{Bioengineering}{10}{3}{334}.
\PrintBackRefs{\CurrentBib}

\bibitem [\protect \citeauthoryear {%
Otazo%
\ \protect \BOthers {.}}{%
Otazo%
\ \protect \BOthers {.}}{%
{\protect \APACyear {2015}}%
}]{%
otazo2015low}
\APACinsertmetastar {%
otazo2015low}%
\begin{APACrefauthors}%
Otazo, R.%
, Candes, E.%
\BCBL {}\ \BBA {} Sodickson, D\BPBI K.%
\end{APACrefauthors}%
\unskip\
\newblock
\APACrefYearMonthDay{2015}{}{}.
\newblock
{\BBOQ}\APACrefatitle {Low-rank plus sparse matrix decomposition for accelerated dynamic {MRI} with separation of background and dynamic components} {Low-rank plus sparse matrix decomposition for accelerated dynamic {MRI} with separation of background and dynamic components}.{\BBCQ}
\newblock
\APACjournalVolNumPages{Magnetic Resonance in Medicine}{73}{3}{1125--1136}.
\PrintBackRefs{\CurrentBib}

\bibitem [\protect \citeauthoryear {%
Otazo%
\ \protect \BOthers {.}}{%
Otazo%
\ \protect \BOthers {.}}{%
{\protect \APACyear {2010}}%
}]{%
otazo2010combination}
\APACinsertmetastar {%
otazo2010combination}%
\begin{APACrefauthors}%
Otazo, R.%
, Kim, D.%
, Axel, L.%
\BCBL {}\ \BBA {} Sodickson, D\BPBI K.%
\end{APACrefauthors}%
\unskip\
\newblock
\APACrefYearMonthDay{2010}{}{}.
\newblock
{\BBOQ}\APACrefatitle {Combination of compressed sensing and parallel imaging for highly accelerated first-pass cardiac perfusion {MRI}} {Combination of compressed sensing and parallel imaging for highly accelerated first-pass cardiac perfusion {MRI}}.{\BBCQ}
\newblock
\APACjournalVolNumPages{Magnetic Resonance in Medicine}{64}{3}{767--776}.
\PrintBackRefs{\CurrentBib}

\bibitem [\protect \citeauthoryear {%
Pedersen%
\ \protect \BOthers {.}}{%
Pedersen%
\ \protect \BOthers {.}}{%
{\protect \APACyear {2009}}%
}]{%
pedersen2009k}
\APACinsertmetastar {%
pedersen2009k}%
\begin{APACrefauthors}%
Pedersen, H.%
, Kozerke, S.%
, Ringgaard, S.%
, Nehrke, K.%
\BCBL {}\ \BBA {} Kim, W\BPBI Y.%
\end{APACrefauthors}%
\unskip\
\newblock
\APACrefYearMonthDay{2009}{}{}.
\newblock
{\BBOQ}\APACrefatitle {k-t {PCA}: temporally constrained k-t {BLAST} reconstruction using principal component analysis} {k-t {PCA}: temporally constrained k-t {BLAST} reconstruction using principal component analysis}.{\BBCQ}
\newblock
\APACjournalVolNumPages{Magnetic Resonance in Medicine}{62}{3}{706--716}.
\PrintBackRefs{\CurrentBib}

\bibitem [\protect \citeauthoryear {%
Pruessmann%
\ \protect \BOthers {.}}{%
Pruessmann%
\ \protect \BOthers {.}}{%
{\protect \APACyear {1999}}%
}]{%
pruessmann1999sense}
\APACinsertmetastar {%
pruessmann1999sense}%
\begin{APACrefauthors}%
Pruessmann, K\BPBI P.%
, Weiger, M.%
, Scheidegger, M\BPBI B.%
\BCBL {}\ \BBA {} Boesiger, P.%
\end{APACrefauthors}%
\unskip\
\newblock
\APACrefYearMonthDay{1999}{}{}.
\newblock
{\BBOQ}\APACrefatitle {{SENSE}: sensitivity encoding for fast {MRI}} {{SENSE}: sensitivity encoding for fast {MRI}}.{\BBCQ}
\newblock
\APACjournalVolNumPages{Magnetic Resonance in Medicine}{42}{5}{952--962}.
\PrintBackRefs{\CurrentBib}

\bibitem [\protect \citeauthoryear {%
Rajiah%
\ \protect \BOthers {.}}{%
Rajiah%
\ \protect \BOthers {.}}{%
{\protect \APACyear {2023}}%
}]{%
rajiah2023cardiac}
\APACinsertmetastar {%
rajiah2023cardiac}%
\begin{APACrefauthors}%
Rajiah, P\BPBI S.%
, Fran{\c{c}}ois, C\BPBI J.%
\BCBL {}\ \BBA {} Leiner, T.%
\end{APACrefauthors}%
\unskip\
\newblock
\APACrefYearMonthDay{2023}{}{}.
\newblock
{\BBOQ}\APACrefatitle {Cardiac {MRI}: state of the art} {Cardiac {MRI}: state of the art}.{\BBCQ}
\newblock
\APACjournalVolNumPages{Radiology}{307}{3}{}.
\newblock
\begin{APACrefURL} \url{http://dx.doi.org/10.1148/radiol.223008} \end{APACrefURL}
\newblock
\APACrefnote{Art. no. e223008}
\PrintBackRefs{\CurrentBib}

\bibitem [\protect \citeauthoryear {%
Ramzi%
\ \protect \BOthers {.}}{%
Ramzi%
\ \protect \BOthers {.}}{%
{\protect \APACyear {2022}}%
}]{%
ramzi2022nc}
\APACinsertmetastar {%
ramzi2022nc}%
\begin{APACrefauthors}%
Ramzi, Z.%
, Chaithya, G\BPBI R.%
, Starck, J\BHBI L.%
\BCBL {}\ \BBA {} Ciuciu, P.%
\end{APACrefauthors}%
\unskip\
\newblock
\APACrefYearMonthDay{2022}{}{}.
\newblock
{\BBOQ}\APACrefatitle {{NC-PDNet}: A density-compensated unrolled network for 2{D} and 3{D} non-Cartesian {MRI} reconstruction} {{NC-PDNet}: A density-compensated unrolled network for 2{D} and 3{D} non-cartesian {MRI} reconstruction}.{\BBCQ}
\newblock
\APACjournalVolNumPages{IEEE Transactions on Medical Imaging}{41}{7}{1625--1638}.
\PrintBackRefs{\CurrentBib}

\bibitem [\protect \citeauthoryear {%
Ronneberger%
\ \protect \BOthers {.}}{%
Ronneberger%
\ \protect \BOthers {.}}{%
{\protect \APACyear {2015}}%
}]{%
ronneberger2015u}
\APACinsertmetastar {%
ronneberger2015u}%
\begin{APACrefauthors}%
Ronneberger, O.%
, Fischer, P.%
\BCBL {}\ \BBA {} Brox, T.%
\end{APACrefauthors}%
\unskip\
\newblock
\APACrefYearMonthDay{2015}{}{}.
\newblock
{\BBOQ}\APACrefatitle {U-net: Convolutional networks for biomedical image segmentation} {U-net: Convolutional networks for biomedical image segmentation}.{\BBCQ}
\newblock
\BIn{} \APACrefbtitle {Proceedings of the 18th International Conference of the Medical Image Computing and Computer-Assisted Intervention ({MICCAI})} {Proceedings of the 18th international conference of the medical image computing and computer-assisted intervention ({MICCAI})}\ (\BPGS\ 234--241).
\PrintBackRefs{\CurrentBib}

\bibitem [\protect \citeauthoryear {%
Roy%
\ \protect \BOthers {.}}{%
Roy%
\ \protect \BOthers {.}}{%
{\protect \APACyear {2022}}%
}]{%
roy2022free}
\APACinsertmetastar {%
roy2022free}%
\begin{APACrefauthors}%
Roy, C\BPBI W.%
, Di~Sopra, L.%
, Whitehead, K\BPBI K.%
, Piccini, D.%
, Yerly, M.%
, Heerfordt, J.%
\BDBL {}Stuber, M.%
\end{APACrefauthors}%
\unskip\
\newblock
\APACrefYearMonthDay{2022}{}{}.
\newblock
{\BBOQ}\APACrefatitle {Free-running cardiac and respiratory motion-resolved 5D whole-heart coronary cardiovascular magnetic resonance angiography in pediatric cardiac patients using ferumoxytol} {Free-running cardiac and respiratory motion-resolved 5d whole-heart coronary cardiovascular magnetic resonance angiography in pediatric cardiac patients using ferumoxytol}.{\BBCQ}
\newblock
\APACjournalVolNumPages{Journal of Cardiovascular Magnetic Resonance}{24}{1}{39}.
\PrintBackRefs{\CurrentBib}

\bibitem [\protect \citeauthoryear {%
Sandino%
\ \protect \BOthers {.}}{%
Sandino%
\ \protect \BOthers {.}}{%
{\protect \APACyear {2021}}%
}]{%
sandino2021accelerating}
\APACinsertmetastar {%
sandino2021accelerating}%
\begin{APACrefauthors}%
Sandino, C\BPBI M.%
, Lai, P.%
, Vasanawala, S\BPBI S.%
\BCBL {}\ \BBA {} Cheng, J\BPBI Y.%
\end{APACrefauthors}%
\unskip\
\newblock
\APACrefYearMonthDay{2021}{}{}.
\newblock
{\BBOQ}\APACrefatitle {Accelerating cardiac cine {MRI} using a deep learning-based {ESPIRiT} reconstruction} {Accelerating cardiac cine {MRI} using a deep learning-based {ESPIRiT} reconstruction}.{\BBCQ}
\newblock
\APACjournalVolNumPages{Magnetic Resonance in Medicine}{85}{1}{152--167}.
\PrintBackRefs{\CurrentBib}

\bibitem [\protect \citeauthoryear {%
Schlemper%
\ \protect \BOthers {.}}{%
Schlemper%
\ \protect \BOthers {.}}{%
{\protect \APACyear {2017}}%
}]{%
schlemper2017deep}
\APACinsertmetastar {%
schlemper2017deep}%
\begin{APACrefauthors}%
Schlemper, J.%
, Caballero, J.%
, Hajnal, J\BPBI V.%
, Price, A\BPBI N.%
\BCBL {}\ \BBA {} Rueckert, D.%
\end{APACrefauthors}%
\unskip\
\newblock
\APACrefYearMonthDay{2017}{}{}.
\newblock
{\BBOQ}\APACrefatitle {A deep cascade of convolutional neural networks for dynamic {MR} image reconstruction} {A deep cascade of convolutional neural networks for dynamic {MR} image reconstruction}.{\BBCQ}
\newblock
\APACjournalVolNumPages{IEEE Transactions on Medical Imaging}{37}{2}{491--503}.
\PrintBackRefs{\CurrentBib}

\bibitem [\protect \citeauthoryear {%
Setser%
\ \protect \BOthers {.}}{%
Setser%
\ \protect \BOthers {.}}{%
{\protect \APACyear {2000}}%
}]{%
Setser2000}
\APACinsertmetastar {%
Setser2000}%
\begin{APACrefauthors}%
Setser, R\BPBI M.%
, Fischer, S\BPBI E.%
\BCBL {}\ \BBA {} Lorenz, C\BPBI H.%
\end{APACrefauthors}%
\unskip\
\newblock
\APACrefYearMonthDay{2000}{Sep}{}.
\newblock
{\BBOQ}\APACrefatitle {{{Q}uantification of left ventricular function with magnetic resonance images acquired in real time}} {{{Q}uantification of left ventricular function with magnetic resonance images acquired in real time}}.{\BBCQ}
\newblock
\APACjournalVolNumPages{Journal of Magnetic Resonance Imaging}{12}{3}{430--438}.
\PrintBackRefs{\CurrentBib}

\bibitem [\protect \citeauthoryear {%
Smith%
\ \BBA {} Nayak%
}{%
Smith%
\ \BBA {} Nayak%
}{%
{\protect \APACyear {2012}}%
}]{%
smith2012reduced}
\APACinsertmetastar {%
smith2012reduced}%
\begin{APACrefauthors}%
Smith, T\BPBI B.%
\BCBT {}\ \BBA {} Nayak, K\BPBI S.%
\end{APACrefauthors}%
\unskip\
\newblock
\APACrefYearMonthDay{2012}{}{}.
\newblock
{\BBOQ}\APACrefatitle {Reduced field of view {MRI} with rapid, {B}1-robust outer volume suppression} {Reduced field of view {MRI} with rapid, {B}1-robust outer volume suppression}.{\BBCQ}
\newblock
\APACjournalVolNumPages{Magnetic Resonance in Medicine}{67}{5}{1316--1323}.
\PrintBackRefs{\CurrentBib}

\bibitem [\protect \citeauthoryear {%
Sun%
\ \protect \BOthers {.}}{%
Sun%
\ \protect \BOthers {.}}{%
{\protect \APACyear {2016}}%
}]{%
sun2016deep}
\APACinsertmetastar {%
sun2016deep}%
\begin{APACrefauthors}%
Sun, J.%
, Li, H.%
\BCBL {}\ \BBA {} Xu, Z.%
\end{APACrefauthors}%
\unskip\
\newblock
\APACrefYearMonthDay{2016}{}{}.
\newblock
{\BBOQ}\APACrefatitle {Deep {ADMM-Net} for compressive sensing {MRI}} {Deep {ADMM-Net} for compressive sensing {MRI}}.{\BBCQ}
\newblock
\APACjournalVolNumPages{Advances in neural information processing systems}{29}{}{}.
\PrintBackRefs{\CurrentBib}

\bibitem [\protect \citeauthoryear {%
Terpstra%
\ \protect \BOthers {.}}{%
Terpstra%
\ \protect \BOthers {.}}{%
{\protect \APACyear {2023}}%
}]{%
terpstra2023accelerated}
\APACinsertmetastar {%
terpstra2023accelerated}%
\begin{APACrefauthors}%
Terpstra, M\BPBI L.%
, Maspero, M.%
, Verhoeff, J\BPBI J\BPBI C.%
\BCBL {}\ \BBA {} Van Den~Berg, C\BPBI A\BPBI T.%
\end{APACrefauthors}%
\unskip\
\newblock
\APACrefYearMonthDay{2023}{}{}.
\newblock
{\BBOQ}\APACrefatitle {Accelerated respiratory-resolved {4D}-{MRI} with separable spatio-temporal neural networks} {Accelerated respiratory-resolved {4D}-{MRI} with separable spatio-temporal neural networks}.{\BBCQ}
\newblock
\APACjournalVolNumPages{Medical Physics}{50}{9}{5331--5342}.
\PrintBackRefs{\CurrentBib}

\bibitem [\protect \citeauthoryear {%
Tian%
\ \protect \BOthers {.}}{%
Tian%
\ \protect \BOthers {.}}{%
{\protect \APACyear {2021}}%
}]{%
tian2021aliasing}
\APACinsertmetastar {%
tian2021aliasing}%
\begin{APACrefauthors}%
Tian, Y.%
, Lim, Y.%
, Zhao, Z.%
, Byrd, D.%
, Narayanan, S.%
\BCBL {}\ \BBA {} Nayak, K\BPBI S.%
\end{APACrefauthors}%
\unskip\
\newblock
\APACrefYearMonthDay{2021}{}{}.
\newblock
{\BBOQ}\APACrefatitle {Aliasing artifact reduction in spiral real-time {MRI}} {Aliasing artifact reduction in spiral real-time {MRI}}.{\BBCQ}
\newblock
\APACjournalVolNumPages{Magnetic Resonance in Medicine}{86}{2}{916--925}.
\PrintBackRefs{\CurrentBib}

\bibitem [\protect \citeauthoryear {%
Tolouee%
\ \protect \BOthers {.}}{%
Tolouee%
\ \protect \BOthers {.}}{%
{\protect \APACyear {2018}}%
}]{%
tolouee2018motion}
\APACinsertmetastar {%
tolouee2018motion}%
\begin{APACrefauthors}%
Tolouee, A.%
, Alirezaie, J.%
\BCBL {}\ \BBA {} Babyn, P.%
\end{APACrefauthors}%
\unskip\
\newblock
\APACrefYearMonthDay{2018}{}{}.
\newblock
{\BBOQ}\APACrefatitle {Motion-compensated data decomposition algorithm to accelerate dynamic cardiac {MRI}} {Motion-compensated data decomposition algorithm to accelerate dynamic cardiac {MRI}}.{\BBCQ}
\newblock
\APACjournalVolNumPages{Magnetic Resonance Materials in Physics, Biology and Medicine}{31}{}{33--47}.
\PrintBackRefs{\CurrentBib}

\bibitem [\protect \citeauthoryear {%
Tsao%
\ \protect \BOthers {.}}{%
Tsao%
\ \protect \BOthers {.}}{%
{\protect \APACyear {2003}}%
}]{%
tsao2003k}
\APACinsertmetastar {%
tsao2003k}%
\begin{APACrefauthors}%
Tsao, J.%
, Boesiger, P.%
\BCBL {}\ \BBA {} Pruessmann, K\BPBI P.%
\end{APACrefauthors}%
\unskip\
\newblock
\APACrefYearMonthDay{2003}{}{}.
\newblock
{\BBOQ}\APACrefatitle {k-t {BLAST} and k-t {SENSE}: dynamic {MRI} with high frame rate exploiting spatiotemporal correlations} {k-t {BLAST} and k-t {SENSE}: dynamic {MRI} with high frame rate exploiting spatiotemporal correlations}.{\BBCQ}
\newblock
\APACjournalVolNumPages{Magnetic Resonance in Medicine}{50}{5}{1031--1042}.
\PrintBackRefs{\CurrentBib}

\bibitem [\protect \citeauthoryear {%
Unterberg-Buchwald%
\ \protect \BOthers {.}}{%
Unterberg-Buchwald%
\ \protect \BOthers {.}}{%
{\protect \APACyear {2014}}%
}]{%
UnterbergBuchwald2014}
\APACinsertmetastar {%
UnterbergBuchwald2014}%
\begin{APACrefauthors}%
Unterberg-Buchwald, C.%
, Fasshauer, M.%
, Sohns, J\BPBI M.%
, Staab, W.%
, Schuster, A.%
, Voit, D.%
\BDBL {}Lotz, J.%
\end{APACrefauthors}%
\unskip\
\newblock
\APACrefYearMonthDay{2014}{Jan}{}.
\newblock
{\BBOQ}\APACrefatitle {Real time cardiac {MRI} and its clinical usefulness in arrhythmias and wall motion abnormalities} {Real time cardiac {MRI} and its clinical usefulness in arrhythmias and wall motion abnormalities}.{\BBCQ}
\newblock
\APACjournalVolNumPages{Journal of Cardiovascular Magnetic Resonance}{16}{}{34}.
\newblock
\begin{APACrefURL} \url{http://dx.doi.org/10.1186/1532-429X-16-S1-P34} \end{APACrefURL}
\PrintBackRefs{\CurrentBib}

\bibitem [\protect \citeauthoryear {%
Van~Amerom%
\ \protect \BOthers {.}}{%
Van~Amerom%
\ \protect \BOthers {.}}{%
{\protect \APACyear {2018}}%
}]{%
van2018fetal}
\APACinsertmetastar {%
van2018fetal}%
\begin{APACrefauthors}%
Van~Amerom, J\BPBI F\BPBI P.%
, Lloyd, D\BPBI F\BPBI A.%
, Price, A\BPBI N.%
, Kuklisova~Murgasova, M.%
, Aljabar, P.%
, Malik, S\BPBI J.%
\BDBL {}others%
\end{APACrefauthors}%
\unskip\
\newblock
\APACrefYearMonthDay{2018}{}{}.
\newblock
{\BBOQ}\APACrefatitle {Fetal cardiac cine imaging using highly accelerated dynamic {MRI} with retrospective motion correction and outlier rejection} {Fetal cardiac cine imaging using highly accelerated dynamic {MRI} with retrospective motion correction and outlier rejection}.{\BBCQ}
\newblock
\APACjournalVolNumPages{Magnetic Resonance in Medicine}{79}{1}{327--338}.
\PrintBackRefs{\CurrentBib}

\bibitem [\protect \citeauthoryear {%
Van~den Bosch%
\ \protect \BOthers {.}}{%
Van~den Bosch%
\ \protect \BOthers {.}}{%
{\protect \APACyear {2008}}%
}]{%
vanDenBosch2008}
\APACinsertmetastar {%
vanDenBosch2008}%
\begin{APACrefauthors}%
Van~den Bosch, M.%
, Daniel, B.%
, Rieke, V.%
, Butts-Pauly, K.%
, Kermit, E.%
\BCBL {}\ \BBA {} Jeffrey, S.%
\end{APACrefauthors}%
\unskip\
\newblock
\APACrefYearMonthDay{2008}{Jan}{}.
\newblock
{\BBOQ}\APACrefatitle {{{M}{R}{I}-guided radiofrequency ablation of breast cancer: Preliminary clinical experience}} {{{M}{R}{I}-guided radiofrequency ablation of breast cancer: Preliminary clinical experience}}.{\BBCQ}
\newblock
\APACjournalVolNumPages{Journal of Magnetic Resonance Imaging}{27}{1}{204--208}.
\PrintBackRefs{\CurrentBib}

\bibitem [\protect \citeauthoryear {%
J.~Wang%
\ \protect \BOthers {.}}{%
J.~Wang%
\ \protect \BOthers {.}}{%
{\protect \APACyear {2021}}%
}]{%
wang2021high}
\APACinsertmetastar {%
wang2021high}%
\begin{APACrefauthors}%
Wang, J.%
, Yang, Y.%
, Weller, D\BPBI S.%
, Zhou, R.%
, Van~Houten, M.%
, Sun, C.%
\BDBL {}Salerno, M.%
\end{APACrefauthors}%
\unskip\
\newblock
\APACrefYearMonthDay{2021}{}{}.
\newblock
{\BBOQ}\APACrefatitle {High spatial resolution spiral first-pass myocardial perfusion imaging with whole-heart coverage at 3{T}} {High spatial resolution spiral first-pass myocardial perfusion imaging with whole-heart coverage at 3{T}}.{\BBCQ}
\newblock
\APACjournalVolNumPages{Magnetic Resonance in Medicine}{86}{2}{648--662}.
\PrintBackRefs{\CurrentBib}

\bibitem [\protect \citeauthoryear {%
X.~Wang%
\ \protect \BOthers {.}}{%
X.~Wang%
\ \protect \BOthers {.}}{%
{\protect \APACyear {2021}}%
}]{%
wang2021fast}
\APACinsertmetastar {%
wang2021fast}%
\begin{APACrefauthors}%
Wang, X.%
, Uecker, M.%
\BCBL {}\ \BBA {} Feng, L.%
\end{APACrefauthors}%
\unskip\
\newblock
\APACrefYearMonthDay{2021}{}{}.
\newblock
{\BBOQ}\APACrefatitle {Fast real-time cardiac {MRI}: A review of current techniques and future directions} {Fast real-time cardiac {MRI}: A review of current techniques and future directions}.{\BBCQ}
\newblock
\APACjournalVolNumPages{Investigative Magnetic Resonance Imaging}{25}{4}{252--265}.
\PrintBackRefs{\CurrentBib}

\bibitem [\protect \citeauthoryear {%
Z.~Wang%
\ \protect \BOthers {.}}{%
Z.~Wang%
\ \protect \BOthers {.}}{%
{\protect \APACyear {2024}}%
}]{%
wang2024deep}
\APACinsertmetastar {%
wang2024deep}%
\begin{APACrefauthors}%
Wang, Z.%
, Xiao, M.%
, Zhou, Y.%
, Wang, C.%
, Wu, N.%
, Li, Y.%
\BDBL {}others%
\end{APACrefauthors}%
\unskip\
\newblock
\APACrefYearMonthDay{2024}{}{}.
\newblock
\APACrefbtitle {Deep separable spatiotemporal learning for fast dynamic cardiac {MRI}.} {Deep separable spatiotemporal learning for fast dynamic cardiac {MRI}.}
\newblock
\begin{APACrefURL} \url{https://arxiv.org/abs/2402.15939} \end{APACrefURL}
\PrintBackRefs{\CurrentBib}

\bibitem [\protect \citeauthoryear {%
Weing{\"a}rtner%
\ \protect \BOthers {.}}{%
Weing{\"a}rtner%
\ \protect \BOthers {.}}{%
{\protect \APACyear {2018}}%
}]{%
weingartner2018feasibility}
\APACinsertmetastar {%
weingartner2018feasibility}%
\begin{APACrefauthors}%
Weing{\"a}rtner, S.%
, Moeller, S.%
\BCBL {}\ \BBA {} Ak{\c{c}}akaya, M.%
\end{APACrefauthors}%
\unskip\
\newblock
\APACrefYearMonthDay{2018}{}{}.
\newblock
{\BBOQ}\APACrefatitle {Feasibility of ultra-high simultaneous multi-slice and in-plane accelerations for cardiac {MRI} using outer volume suppression and leakage-blocking reconstruction} {Feasibility of ultra-high simultaneous multi-slice and in-plane accelerations for cardiac {MRI} using outer volume suppression and leakage-blocking reconstruction}.{\BBCQ}
\newblock
\BIn{} \APACrefbtitle {Proceedings of the 27th International Society for Magnetic Resonance in Medicine ({ISMRM}).} {Proceedings of the 27th international society for magnetic resonance in medicine ({ISMRM}).}
\newblock
\begin{APACrefURL} \url{https://archive.ismrm.org/2018/0359.html} \end{APACrefURL}
\newblock
\APACrefnote{Abstract No. 0359}
\PrintBackRefs{\CurrentBib}

\bibitem [\protect \citeauthoryear {%
Yaman%
\ \protect \BOthers {.}}{%
Yaman%
\ \protect \BOthers {.}}{%
{\protect \APACyear {2022}}%
}]{%
yaman2022multi}
\APACinsertmetastar {%
yaman2022multi}%
\begin{APACrefauthors}%
Yaman, B.%
, Gu, H.%
, Hosseini, S\BPBI A\BPBI H.%
, Demirel, {\"O}\BPBI B.%
, Moeller, S.%
, Ellermann, J.%
\BDBL {}Ak{\c{c}}akaya, M.%
\end{APACrefauthors}%
\unskip\
\newblock
\APACrefYearMonthDay{2022}{Dec}{}.
\newblock
{\BBOQ}\APACrefatitle {{{M}ulti-mask self-supervised learning for physics-guided neural networks in highly accelerated magnetic resonance imaging}} {{{M}ulti-mask self-supervised learning for physics-guided neural networks in highly accelerated magnetic resonance imaging}}.{\BBCQ}
\newblock
\APACjournalVolNumPages{Nuclear Magnetic Resonance in Biomedicine}{35}{12}{}.
\newblock
\APACrefnote{Art. no. e4798}
\PrintBackRefs{\CurrentBib}

\bibitem [\protect \citeauthoryear {%
Yaman%
\ \protect \BOthers {.}}{%
Yaman%
\ \protect \BOthers {.}}{%
{\protect \APACyear {2020}}%
{\protect \APACexlab {{\protect \BCnt {1}}}}}]{%
yaman2020self}
\APACinsertmetastar {%
yaman2020self}%
\begin{APACrefauthors}%
Yaman, B.%
, Hosseini, S\BPBI A\BPBI H.%
, Moeller, S.%
, Ellermann, J.%
, U{\u{g}}urbil, K.%
\BCBL {}\ \BBA {} Ak{\c{c}}akaya, M.%
\end{APACrefauthors}%
\unskip\
\newblock
\APACrefYearMonthDay{2020{\protect \BCnt {1}}}{Dec}{}.
\newblock
{\BBOQ}\APACrefatitle {{{S}elf-supervised learning of physics-guided reconstruction neural networks without fully sampled reference data}} {{{S}elf-supervised learning of physics-guided reconstruction neural networks without fully sampled reference data}}.{\BBCQ}
\newblock
\APACjournalVolNumPages{Magnetic Resonance in Medicine}{84}{6}{3172--3191}.
\PrintBackRefs{\CurrentBib}

\bibitem [\protect \citeauthoryear {%
Yaman%
\ \protect \BOthers {.}}{%
Yaman%
\ \protect \BOthers {.}}{%
{\protect \APACyear {2020}}%
{\protect \APACexlab {{\protect \BCnt {2}}}}}]{%
yaman2020selfconf}
\APACinsertmetastar {%
yaman2020selfconf}%
\begin{APACrefauthors}%
Yaman, B.%
, Hosseini, S\BPBI A\BPBI H.%
, Moeller, S.%
, Ellermann, J.%
, U{\u{g}}urbil, K.%
\BCBL {}\ \BBA {} Ak{\c{c}}akaya, M.%
\end{APACrefauthors}%
\unskip\
\newblock
\APACrefYearMonthDay{2020{\protect \BCnt {2}}}{}{}.
\newblock
{\BBOQ}\APACrefatitle {Self-supervised physics-based deep learning {MRI} reconstruction without fully-sampled data} {Self-supervised physics-based deep learning {MRI} reconstruction without fully-sampled data}.{\BBCQ}
\newblock
\BIn{} \APACrefbtitle {Proceedings of the IEEE 17th International Symposium on Biomedical Imaging ({ISBI})} {Proceedings of the ieee 17th international symposium on biomedical imaging ({ISBI})}\ (\BPGS\ 921--925).
\PrintBackRefs{\CurrentBib}

\bibitem [\protect \citeauthoryear {%
Yang%
\ \protect \BOthers {.}}{%
Yang%
\ \protect \BOthers {.}}{%
{\protect \APACyear {2018}}%
}]{%
yang2018reduced}
\APACinsertmetastar {%
yang2018reduced}%
\begin{APACrefauthors}%
Yang, Y.%
, Zhao, L.%
, Chen, X.%
, Shaw, P\BPBI W.%
, Gonzalez, J\BPBI A.%
, Epstein, F\BPBI H.%
\BDBL {}Salerno, M.%
\end{APACrefauthors}%
\unskip\
\newblock
\APACrefYearMonthDay{2018}{}{}.
\newblock
{\BBOQ}\APACrefatitle {Reduced field of view single-shot spiral perfusion imaging} {Reduced field of view single-shot spiral perfusion imaging}.{\BBCQ}
\newblock
\APACjournalVolNumPages{Magnetic Resonance in Medicine}{79}{1}{208--216}.
\PrintBackRefs{\CurrentBib}

\bibitem [\protect \citeauthoryear {%
S.~Zhang%
\ \protect \BOthers {.}}{%
S.~Zhang%
\ \protect \BOthers {.}}{%
{\protect \APACyear {2014}}%
}]{%
zhang2014real}
\APACinsertmetastar {%
zhang2014real}%
\begin{APACrefauthors}%
Zhang, S.%
, Joseph, A\BPBI A.%
, Voit, D.%
, Schaetz, S.%
, Merboldt, K\BHBI D.%
, Unterberg-Buchwald, C.%
\BDBL {}Frahm, J.%
\end{APACrefauthors}%
\unskip\
\newblock
\APACrefYearMonthDay{2014}{}{}.
\newblock
{\BBOQ}\APACrefatitle {Real-time magnetic resonance imaging of cardiac function and flow—recent progress} {Real-time magnetic resonance imaging of cardiac function and flow—recent progress}.{\BBCQ}
\newblock
\APACjournalVolNumPages{Quantitative Imaging in Medicine and Surgery}{4}{5}{313}.
\PrintBackRefs{\CurrentBib}

\bibitem [\protect \citeauthoryear {%
Y.~Zhang%
\ \protect \BOthers {.}}{%
Y.~Zhang%
\ \protect \BOthers {.}}{%
{\protect \APACyear {2024}}%
}]{%
zhang2024direct}
\APACinsertmetastar {%
zhang2024direct}%
\begin{APACrefauthors}%
Zhang, Y.%
, Stolt-Ans{\'o}, N.%
, Pan, J.%
, Huang, W.%
, Hammernik, K.%
\BCBL {}\ \BBA {} Rueckert, D.%
\end{APACrefauthors}%
\unskip\
\newblock
\APACrefYearMonthDay{2024}{}{}.
\newblock
{\BBOQ}\APACrefatitle {Direct cardiac segmentation from undersampled k-space using transformers} {Direct cardiac segmentation from undersampled k-space using transformers}.{\BBCQ}
\newblock
\BIn{} \APACrefbtitle {Proceedings of the IEEE 21st International Symposium on Biomedical Imaging ({ISBI})} {Proceedings of the ieee 21st international symposium on biomedical imaging ({ISBI})}\ (\BPGS\ 1--4).
\PrintBackRefs{\CurrentBib}

\bibitem [\protect \citeauthoryear {%
Zhu%
\ \protect \BOthers {.}}{%
Zhu%
\ \protect \BOthers {.}}{%
{\protect \APACyear {2023}}%
}]{%
zhu2023physics}
\APACinsertmetastar {%
zhu2023physics}%
\begin{APACrefauthors}%
Zhu, Y.%
, Cheng, J.%
, Cui, Z\BHBI X.%
, Zhu, Q.%
, Ying, L.%
\BCBL {}\ \BBA {} Liang, D.%
\end{APACrefauthors}%
\unskip\
\newblock
\APACrefYearMonthDay{2023}{}{}.
\newblock
{\BBOQ}\APACrefatitle {Physics-driven deep learning methods for fast quantitative magnetic resonance imaging: Performance improvements through integration with deep neural networks} {Physics-driven deep learning methods for fast quantitative magnetic resonance imaging: Performance improvements through integration with deep neural networks}.{\BBCQ}
\newblock
\APACjournalVolNumPages{IEEE Signal Processing Magazine}{40}{2}{116--128}.
\PrintBackRefs{\CurrentBib}

\bibitem [\protect \citeauthoryear {%
Zu%
\ \protect \BOthers {.}}{%
Zu%
\ \protect \BOthers {.}}{%
{\protect \APACyear {2013}}%
}]{%
Zu2013}
\APACinsertmetastar {%
Zu2013}%
\begin{APACrefauthors}%
Zu, Y.%
, Narayanan, S\BPBI S.%
, Kim, Y\BHBI C.%
, Nayak, K.%
, Bronson-Lowe, C.%
, Villegas, B.%
\BDBL {}Sinha, U\BPBI K.%
\end{APACrefauthors}%
\unskip\
\newblock
\APACrefYearMonthDay{2013}{Dec}{}.
\newblock
{\BBOQ}\APACrefatitle {{{E}valuation of swallow function after tongue cancer treatment using real-time magnetic resonance imaging: a pilot study}} {{{E}valuation of swallow function after tongue cancer treatment using real-time magnetic resonance imaging: a pilot study}}.{\BBCQ}
\newblock
\APACjournalVolNumPages{JAMA Otolaryngology Head Neck Surgery}{139}{12}{1312--1319}.
\PrintBackRefs{\CurrentBib}

\end{thebibliography}

\end{document}